\documentclass{article}[12pts]
\usepackage[]{amsmath,amssymb}
\usepackage{epic}
\usepackage{graphics,epsfig}
\usepackage{color}
\usepackage{amsfonts}
\newcommand{\ud}[1]{\underline{#1}}
\newcommand{\udd}[1]{\underline{\underline{#1}}}
\newcommand{\ov}[1]{\overline{#1}}

\def \otimesdot {\stackrel{\cdot}{\otimes}}
\def\ie{{\rm i.e.,\/}\ }
\def\etc{{\rm etc.\/}\ }
\let\sect=\section
\def\section{\newpage\sect}

\def\text#1{\mbox{\rm #1\ }}

\def\ie{{\rm i.e.,\/}\ }
\def\etc{{\rm etc.\/}\ }

\def\one{\mbox{\rm 1}\hskip-2.8pt \mbox{\rm l}}
\def \otimesdot {\stackrel{\cdot}{\otimes}}
\newcommand{\ZZ}{\mathbb{Z}}

\newcommand{\CC}{\mathbb{C}}

\title{
Determination of quantum symmetries for higher\\  $ADE$ systems 
from the modular $T$ matrix  
\vspace{0.8cm}}

\author{R. Coquereaux${}^{1}$ \thanks{~Email:
Robert.Coquereaux@cpt.univ-mrs.fr}
$\;$,
                  G. Schieber${}^{1,}$  ${}^2$ \thanks{~Email:
schieber@if.ufrj.br} \\
\\
${}^1$ {\it Centre de Physique Th\'eorique and CIRM} \\
                 {\it Campus de Luminy}           \\
                 {\it F-13288 Marseille - France}            \\
\\
${}^2$ {\it Instituto de F\'{\i}sica, Universidade Federal do Rio de
Janeiro} \\
                 {\it Ilha do Fund\~ao, Caixa Postal 68528}\\
                 {\it 21941-972, Rio de Janeiro, Brasil}\\
}

\date{}
\begin{document}
\thispagestyle{empty}
\begin{titlepage}
\maketitle
%\vfill
\abstract{We show that the Ocneanu algebra of quantum symmetries, for
an $ADE$ diagram (or for  higher Coxeter-Dynkin systems, like
the Di Francesco - Zuber system) can be, in most cases, deduced from the structure of
the modular $T$ matrix in the $A$ series.
We recover in this way the (known) quantum symmetries of $su(2)$
diagrams and illustrate our method by studying those
associated with the three genuine exceptional diagrams of type
$su(3)$, namely ${\cal E}_5$,  ${\cal E}_9$ and  ${\cal E}_{21}$.
This also provides the shortest way to the determination of twisted 
partition functions in boundary conformal field theory with
defect lines.}

\vspace{3.cm}

\vspace{1.0cm}

%\noindent Anonymous ftp: cpt.univ-mrs.fr

\vspace{1. cm}

\noindent {\tt  hep-th/0203242}\\
\noindent CPT-2002/P.4336 \\

\vspace*{0.3 cm}

\end{titlepage}

%%%%%%%%%%%%%%%%%%%%%%%%%%%%%%%%%%%%%%%%%%%%%%%%%%%%%%%%%%%%%%%%%%%%%%%%%
%%%%%%%%%%%%%%%%%%%%%%%%%%%%%%%%%%%%%%%%%%%%%%%%%%%%%%%%%%%%%%%%%%%%%%%%%

\section{Introduction}

This article  provides a simple tool for the determination, in most cases,  of
the algebra of quantum symmetries associated with $ADE$ Dynkin
diagrams (considered as quantum $su(2)$ objects) or with their
generalizations to higher systems (Di Francesco - Zuber diagrams in
the case of $su(3)$).

Although a precise general definition of extended Coxeter-Dynkin systems is
still lacking, the known examples
always contain
a ``principal'' series (the ${\cal A}$ series) and a finite number of
``genuine exceptional'' cases \cite{Ocneanu:MSRI}.
The other diagrams of the system are obtained  as orbifolds of the
genuine diagrams (exceptional or not) and as twists or conjugates
(sometimes both) of
the genuine diagrams and of their orbifolds.
In the case of $su(2)$ (the usual $ADE$ system), we have the principal
$A$ series, and the two genuine
exceptional cases $E_6$ and $E_8$; the $D_{2n}$ diagrams are orbifolds of
the $A_{4n-3}$ diagrams; the  $D_{2n+1}$ diagrams are orbifolds of the 
$A_{4n-1}$ diagrams
and $E_7$ is a twist of the $D_{10}$ diagram (itself an orbifold of
$A_{17}$).
In the case of $su(3)$ (the Di Francesco - Zuber system, slightly amended by 
A.Ocneanu in \cite{Ocneanu:Bariloche}),
we have the principal series ${\cal A}_k$, and three genuine exceptional
diagrams: ${\cal E}_5$,  ${\cal E}_9$ and  ${\cal E}_{21}$; the others 
(in particular the  other four exceptionals) of the system
are obtained from these genuine diagrams by orbifolding, twisting and
conjugating.

In some cases, the vector space spanned by the vertices of a given diagram
$G$ admits ``self-fusion'' \cite{Pasquier:alg}, \cite{Pasquier}, \ie it 
possesses an associative algebra structure with
positive integral structure constants (like $A_n$, $D_{2n}$, $E_6$
and $E_8$ for the $su(2)$ system).  Sometimes
it does not (like $D_{2n+1}$ and $E_7$). In all cases, this vector
space is a module over the associative algebra of
the particular diagram $A$ of the ${\cal A}$ series which has the same
Coxeter number (whose definition has to be suitably generalized for
the higher systems).

The  ${\cal A}$ series is always modular: one can define a
representation of $SL(2,\ZZ)$ on the vector space of every diagram
of this class (actually this representation
factors to a finite group, but we shall not need this information
here). The standard generators of this group are called $S$
and $T$. The vector space of the chosen diagram comes
with a particular basis, where the basis vectors  are associated with  graph vertices.
The operator $T$ is diagonal on the vertices.

Take $G$ a diagram and $A$ the corresponding member of the ${\cal A}$
series.  Being a module over the algebra of $A$, there exist
induction-restriction maps between $G$ and $A$ and one can try to
define an action of $SL(2,\ZZ)$ on the vector space of $G$, in a way
that would be compatible with those maps; this is not necessarily
possible.  In plain terms: suppose that the vertex $\sigma$ of $G$
appears both in the branching rules (restriction map from $A$ to $G$)
of vertices $\tau_p$ and $\tau_q$ of $A$; one could think of defining
the value of the modular generator $T$ on $\sigma$ either as
$T(\tau_p)$ or as $T(\tau_q)$, but this is ambiguous, unless these two
values are equal.  In general, there is only a subset $J$ of the
vertices of $G$ for which $T$ can be defined: a vertex $\sigma$ will
belong to this subset whenever $T$ is constant along the vertices of
$A$ whose restriction to $G$ contains $\sigma$. 

 Following Ocneanu
\cite{Ocneanu:paths}, to every diagram $G$ (with or without
self-fusion) belonging to a Coxeter-Dynkin system, one can associate a
bialgebra $\mathcal{B}G$. This bialgebra should be, technically, a weak Hopf
algebra -- or quantum groupoid-- and we have checked this in a few
cases, but we are not aware of any general proof (see our comments in
the final section). By using a particular scalar
product, one can trade the comultiplication for a multiplication and
think that $\mathcal{B}G$ is a di-algebra rather than a bialgebra. 
There are two --usually distinct -- block decompositions for this
di-algebra.  Blocks of the first type are labelled by points of a
diagram $A$ (the member of the ${\cal A}$ series that has same Coxeter
number as $G$).  Blocks of the second type are labelled by points of
another diagram that we call $Oc(G)$.  The two sets of orthogonal
projectors associated with these two block decompositions can be
multiplied with either of these two associative multiplications and
this allows one to define associative algebra structures on the vector
spaces spanned by the vertices of the two graphs $A$ and
$Oc(G)$.  We denote these algebras by the same symbol as the graphs
themselves.  In the particular case where $G$ is a member of the
$\mathcal{A}$ series, these algebras co\" \i ncide.  In all cases,
$A$ is an algebra with a single generator and it is commutative. 
$Oc(G)$, also called ``algebra of quantum symmetries of $G$'', is in
general an algebra with two generators (only one if $G=A$) and it is
not always commutative. 

   In the cases where $Oc(G)$ is commutative, we
observe that this algebra of quantum symmetries can be written in
terms of a tensor product of appropriate graph algebras, but the
tensor product should be taken above some subalgebra determined by the
modular properties of the graph $G$ and we refer to section 3 for a
discussion of the several $ADE$ cases.  Paradoxically, the simplest
cases (besides the $A_{n}$) are those where the diagram $G$ is an
exceptional diagram equal to $E_6$ or $E_8$ (notice that $E_7$ does
not enjoy self-fusion); in those simple cases $Oc(G)$ is isomorphic
with $G\otimes_{J} G$, where $J$ is the particular
subalgebra of the graph algebra of $G$ whose determination (using
modular considerations) was sketched previously.  The tensor product
sign, taken ``above $J$'', means that we identify $au \otimes b$ and
$a \otimes ub$ whenever $u \in J \subset G$.  When $Oc(G)$ is not
commutative, the method is not fully satisfactory, as we shall see.

 The structure of our article is as follows.  The first section reminds
the reader several useful (but not necessarily widely known) facts
about graph algebras and their quantum symmetries.  It also precises
our notations.  The reader already familiar with quantum symmetries of
graphs may skip this part.  In the second section, we consider the
$su(2)$ Coxeter-Dynkin system, \ie the usual $ADE$ diagrams.  For
every one of them we simply recover the structure of $Oc(G)$ by our
method (which is not fully satisfactory for $D_{2n}$, since the
algebra of quantum symmetries of the later is non commutative).  We
give more details on the $E_6$ case because it is both nice and
pedagogical.  In the third section, we move to the $su(3)$
Coxeter-Dynkin system.  After some generalities on these Di Francesco
- Zuber graphs and a short description of the cases associated with
diagrams of ${\cal A}$ type (which are relatively trivial), we study,
in details, also because it is simple enough to be pedagogical, the
quantum symmetries of the diagram ${\cal E}_5$ (the David star), which
is one of the three genuine exceptional cases and is a module over
${\cal A}_5$.  The technique being now clear, we list only the results
for the other two genuine exceptional diagrams ${\cal E}_9$ and ${\cal
E}_{21}$, \ie we give their induction-restriction graphs, the values
of the modular operator $T$ and, for ${\cal E}_{5}$ and ${\cal
E}_{21}$, the structure of their Ocneanu graph.  To every point of
such a graph, one may associate a ``toric matrix''
\cite{Coque:Qtetra}, \cite{Ocneanu:paths}, or, equivalently, a twisted
partition function in boundary conformal field theory with defect
lines \cite{PetZub:Oc}; we also give their explicit expressions for
the studied $su(3)$ cases, at least those associated with the
so-called ambichiral points (to keep the size of this paper
reasonable).  The list of Di Francesco - Zuber graphs being quite
long, we stop at this point, but all the other associated Ocneanu
graphs should be obtained by proper generalizations of the study made
for $su(2)$; the details can, admittedly, be quite intricate, in
particular for those graphs for which $Oc(G)$ is not commutative.

Many topics discussed in the present paper are already known to
experts. We believe however that a systematic discussion of the
correspondence between the eigenvalues of the $T$ operators and the
determination of quantum symmetries is not available elsewhere. Our
explicit results concerning the Ocneanu graphs of several
exceptional diagrams of the $su(3)$ system seem also to be new, and,
we hope, of interest for the reader.

\section{About Coxeter-Dynkin graph algebras and
their quantum symmetries }

\subsection{Generalities}

To a diagram $G$ belonging to a (possibly higher) Coxeter-Dynkin
system, one can associate \cite{Ocneanu:paths} a bialgebra ${\cal B}(G)$
that we call Ocneanu-Racah-Wigner bialgebra (the precise
definition of this bialgebra uses the notion of essential paths on the
graph $G$: see our discussion in the Appendix). 
According to A. Ocneanu (unpublished), this object, also called
``algebra of double triangles'', is a semi-simple weak Hopf algebra
(or quantum groupoid) --- see \cite{Sz}, \cite{Wainerman-1}, for
general properties of quantum groupoids.  We shall not use it
explicitly in our paper and it is enough to say that, as a bialgebra, it
possesses two associative algebra structures (say ``composition
$\circ$'' and ``convolution $\star$''), for which the underlying
vector space can be block diagonalized (\ie decomposed as a sum of
matrix algebras) in two different ways.  Diagonalization of the
convolution product is encoded by a finite dimensional algebra $Oc(G)$
called ``algebra of quantum symmetries''.  As a vector space, $Oc(G)$
contains one linear generator for every single block of $({\cal B}(G),
\star)$.  As an algebra, it has a unit called $\underline 0$ and two
generators called ${\underline 1}_L$ and ${\underline 1}_R$, which,
when $G$ is a member of an ${\cal A}$ series, coincide.  Like the
graph algebra of $G$ (when it exists), the algebra $Oc(G)$ comes with
a preferred basis.  Even when the vector space of $G$ does not admit
self-fusion, so that it is only a module over the corresponding ${\cal
A}$, the associated object $Oc(G)$ is always both an associative
algebra and a bimodule over ${\cal A} \otimes {\cal A}$.  This last
structure is encoded by a set of matrices that we call ``toric
matrices''; there is one such matrix for every point of the Ocneanu
graph.  The multiplicative structure of $Oc(G)$ is fully determined by
the two Cayley graphs of multiplication by the generators; the union
of these two graphs is called the Ocneanu graph of $G$ and is denoted
by the same symbol.  In most cases, $Oc(G)$ is isomorphic with a
tensor product -- over a particular subalgebra $J$ -- of two
associative and commutative algebras; we write $\otimesdot \equiv
\otimes_{J}$ this tensor product; in these cases, $Oc(G)$ is
commutative.  When it is not commutative (the case of
$D_{2n}$ for the $su(2)$ system), one has also to add some
matrix algebra component to this tensor product, in order to take the
non-commutativity into account (see \cite{CoqueGil:ADE} for explicit
formulas for $D_{2n}$ cases).  The two generators of
$Oc(G)$ read ${\underline 1}_L = 1\otimesdot 0$ and ${\underline 1}_R
= 0 \otimesdot 1$.  Their algebraic span are respectively the ``left
chiral'' and ``right chiral'' parts.  The intersection of chiral
subalgebras is called ``ambichiral'' and the vector space spanned by
those (preferred) linear generators which belong to none of the chiral
parts is called ``the supplementary part''.  All these structures lead
to ``nimreps'' (non-negative integer valued matrix representations) of
certain algebras \cite{PetZub2}.

From the point of view of Conformal Field Theory, we are interested in
partition functions on a torus with defect lines.  When there are no
defects these partition functions are modular invariant; this is
usually not so in the presence of defects.  In all cases, they are
sesquilinear forms with non negative integer entries defined on the
vector space spanned by the characters of an affine Lie algebra
$\widehat{\cal G}$.  Here we forget this interpretation and replace
these characters by vertices of a diagram of type ${\cal A}$. 
Partition functions are therefore square matrices indexed by these
vertices.  It was recognized more than seven years ago
by A. Ocneanu (published reference is
\cite{Ocneanu:paths}) that ``the'' modular invariant of
Capelli-Itzykson-Zuber \cite{CapItzZub}, \cite{Ostrik},
\cite{Slodowy}, for a given $ADE$ diagram $G$, was given by the toric
matrix $W_0$ associated with the origin $\underline 0$ of the graph
$Oc(G)$.  To see an example of how all this works, the reader may look
at \cite{Coque:Qtetra}, where toric matrices $W_x$ associated with the
twelve points $x$ of the graph $Oc(E_6)$ are calculated.  In
\cite{PetZub:bcft} it was shown (among other things) that to the other
points -- other than the origin -- of a graph $Oc(G)$ can be
associated partition functions in boundary conformal field theory
(BCFT) with one defect line; these functions are not modular
invariant.  More general toric matrices (or partition functions)
$W_{x,y}$, associated to BCFT with two defect lines, were also
introduced in the same paper (note: $W_x \equiv W_{x,\underline{0}}$). 
Fully explicit expressions for the twisted partition functions $W_x$
are given in \cite{CoqueGil:ADE}, for all $ADE$ cases, by using the
formalism introduced in \cite{Coque:Qtetra}.  This was done
independently of the work \cite{PetZub:Oc}.  It should probably be
stressed that all these expressions were already obtained (but
unpublished) almost eight years ago by A. Ocneanu himself.

The direct determination of the algebra $Oc(G)$, with the definition
provided by A. Ocneanu, is not an easy task and
the associated graphs  are only known (published) for the $su(2)$
Coxeter-Dynkin system.
One of the purposes of \cite{Coque:Qtetra} and
\cite{CoqueGil:ADE}, besides the calculation of the toric matrices,
     was actually to give an algebraic construction  providing a
realization of the {\sl algebra} $Oc(G)$ in terms of graph
algebras associated with appropriated Dynkin diagrams.
In the simple cases (paradoxically, for Dynkin diagrams,  besides
the $A_{n}$ themselves, the ``simple'' cases happen to be those
where $G$ is an exceptional diagram
equal to $E_{6}$ or $E_{8}$), the algebra of quantum symmetries is isomorphic
with $G\otimes_{J} G$, where $J$ is a particular subalgebra of the
algebra of $G$ (we refer to \cite{CoqueGil:ADE} for a
discussion of all $ADE$ cases).
The tensor product sign, taken ``above $J$'',  means
that we identify $au \otimes b$ and $a \otimes ub$ whenever $u \in J
\subset G$. In the last quoted reference, the Ocneanu {\sl graphs},
determined by Ocneanu himself, had to be taken as an input. This was a
weak point in our approach.

For the $su(2)$ Dynkin system, \ie for $ADE$ diagrams, 
one purpose of the present article is to show that the structure of
$Oc(G)$, can be, in most cases,
determined from the eigenvalues of the modular $T$ matrix in
the Hurwitz-Verlinde
representation \cite{Baki}, \cite{Hurwitz}, \cite{Verlinde}, associated with the
graph algebra of $A_{n}$.
The method is general but its implementation depends about the type
of diagram considered, \ie whether it is
a member of the ${\cal A}$ series, a genuine exceptional, or if it
is obtained as an orbifold or by twisting.
In any case, one has first to select a particular subspace $J$ 
by using the list of eigenvalues of the modular
operator $T$ acting on the vertices belonging to the corresponding 
${\cal A}$ diagram.
In the case of $E_6$, for instance, the subset $J$, obtained as
explained in the introduction,
by using a modular constraint on the induction-restriction rules
coming from the $A_{11}$ action,
is isomorphic with an $A_3$ subalgebra of $E_6$ and the Ocneanu algebra
$Oc(E_6)$ is recognized as $E_6 \otimesdot_{A_3} E_6$.  
Warning: everywhere in this paper, the symbol denoting
the diagram also denotes its corresponding associative graph algebra,
when it exists; it never refers to the corresponding Lie algebra 
with the same name (for the higher Coxeter-Dynkin systems, this 
would not even be an algebra in the usual sense!).
The analysis of the $D_{2n}$ cases, where $Oc(G)$ is not commutative,
is more subtle.

For the $su(3)$ system, a direct diagonalization of the convolution
law $\star$ of the bialgebra ${\cal B} (G)$ was
never performed explicitly (or maybe by A. Ocneanu, but this information is not
available), and the algebras $Oc(G)$ -- or their Cayley graphs --
have never been calculated (published) or even properly defined;
therefore our method, which can indeed be generalized in a
straightforward manner to
this more general setting, has a conjectural flavor since
we do not compare our results with those that would be obtained by
a direct approach. 
Nevertheless, we have checked, in the case of exceptional
graphs of $su(3)$ type, that partition functions (toric matrices)
associated with the
origin of ``our'' Ocneanu graphs indeed coincide with the modular
invariant partition functions calculated by \cite{Gannon}
and that expected sum rules also hold (non trivial equalities between
two sums of
squares coming from the diagonalization of the two associative structures
for a given bialgebra).
We obtain also, as a by - product,
the list of twisted partition functions corresponding to a given
diagram $G$ (there are $24$ of them for the exceptional ${\cal
E}_5$ case of the $su(3)$ system).

\subsection{Useful formulae and notations}

For Dynkin diagrams, \ie the $su(2)$ system, $\kappa$ is the (dual) Coxeter
number of the diagram itself. It can be defined, without any
reference to the theory of Lie algebras, from the norm $\beta$ of the graph
(biggest eigenvalue of the adjacency matrix):
$\beta$ is  equal to $2 \cos \left( \frac{ \pi}{\kappa} \right) $.
Note that $1 < \beta < 2$ (see also \cite{Jones:book}).
For Di Francesco - Zuber graphs, \ie the $su(3)$ system, the norm
$\beta$ is equal to $1+ 2 \cos \left( \frac{2 \pi}{\kappa} \right)$.
Note that $2 < \beta < 3$.
This again defines the integer $\kappa$. We call it the ``generalized Coxeter
number of the graph'' or ``altitude'' (like in \cite{DiFZub}).
We also define $q  =  exp{\frac{i \pi}{\kappa}}$, so that
$q^{2\kappa} = 1$.
Another integer $h$ characterizes the system of diagrams.  For Dynkin diagrams,
$h=2$, the (dual) Coxeter number of $su(2)$.
For Di Francesco - Zuber graphs, $h = 3$, the (dual) Coxeter number of
$su(3)$.

The level $k$ of a {\sl diagram} is defined by the relation  $k
 =  \kappa -h$.
Notation for graphs: we keep the standard notation for usual Dynkin diagrams,
with subscript referring to the
number of vertices, \ie the rank of the corresponding Lie algebra.
However, for consistency
with the notation used for higher Coxeter-Dynkin systems, it would be
better for this subscript
to refer to the level $k$ or to the altitude $\kappa$.
We may use both notations, but with script capitals in the later case,
for instance (Dynkin diagrams): ${\cal A}_{\kappa-2} = A_{\kappa
-1}$, ${\cal E}_{10} = E_6$,  ${\cal E}_{16} = E_7$,  ${\cal E}_{28}
= E_8$.
In the case of the Di Francesco - Zuber system of graphs, our subscript
will always refer to the level. Since $h=3$ for
$su(3)$, we have $k = \kappa -3$ for all diagrams of this family.
The reader should be warned that this notation is not
universally accepted, and some authors may prefer to use the
altitude (as an upper index) rather than the level.
For instance, the graphs that we call ${\cal E}_5$, ${\cal E}_9$ and
${\cal E}_{21}$
(like in \cite{Ocneanu:Bariloche})  were called respectively
${\cal E}^{(8)}$, ${\cal E}^{(12)}_{2}$ and ${\cal E}^{(24)}$ in \cite{DiFZub}.

In the case of $su(N)$, there are $N-1$
fundamental representations $f$, and therefore $N-1$ graphs $G^{f}$
(see \cite{DiFZub}),
representing tensor multiplication of irreps by $f$. Since we shall
work only with
$su(2)$ or $su(3)$, we need only one graph. In the case of $su(2)$,
this is clear. In the case of $su(3)$,
this graph is associated with one fundamental irrep (say $3$), the other
graph associated with its conjugate (say $\overline 3$) is obtained by
reversing all the arrows; adjacency matrices corresponding to the
fundamental and to its conjugate are denoted by $G$ and by its transpose $G^T$.

For a diagram of type $su(N)$, the graph  algebra, when it
exists, is faithfully represented
(regular representation) by  $r\times r$ matrices $G_{a}$. In all cases, $G_{0}$ is
the identity matrix and $G_{1}$ is the adjacency matrix.
We denote by $r$ the number of vertices of the diagram $G$.
The $r$ linear generators $\sigma_a$ of $G$, with dual Coxeter number
(or altitude) $\kappa$ are  then represented by $r$ commuting  matrices $G_a$.

In the particular case where $G$ is a member of the $A$ system, the generators
will be called $\tau_i$ and
the corresponding matrices will be called $N_i$.
For a diagram  of type $A$ belonging to a given $su(N)$ system,
writing down matrices $N_{0}$ (identity) and $N_{1}$ (adjacency
matrix) is immediate, and
there are always simple recurrence
formulae that allow one to compute the matrices $N_i$ for all vertices of the
$A$ system in terms of $N_{0}$ and $N_{1}$ (thought as the basic
representation).
These standard recurrence formulae can be obtained for instance by
making products
of Young frames (see later sections for $su(2)$ and $su(3)$).

The module property (external multiplication) of the vector space
associated with a diagram $G$,
of level $k$ and possessing $r$ vertices,  with respect to an action of the 
corresponding algebra ${\cal
A}_{k}$ is encoded by a set of $s$  matrices $F_{i}, {i=0 \ldots
s-1}$, of dimension $r\times r$, sometimes called ``fused graph
matrices'' (a misleading terminology!): $\tau_{i} \sigma_{a} = \sum_{b} (F_{i})_{ab}
\sigma_{b}$.  The number $s$ of vertices of ${\cal A}_{k}$  depends on the
system: for Dynkin diagrams (${\cal A}_{k} = A_{k+1}$), $s = k+1$;
for Di Francesco - Zuber
graphs, $s = (k+1)(k+2)/2$.  Matrix $F_{0}$ is the identity and
matrix $F_{1}=G_1$ is also the adjacency matrix of $G$.
The other $F$ matrices are determined by imposing that they should
obey the same recurrence
relation as the $N$ matrices; this ensures compatibility with left
multiplication by the algebra ${\cal A}_{k}$.
The sets of matrices $F_{i}$, $N_{i}$ and $G_{a}$ of course coincide when
$G$ is a diagram of type $A$.
The $r$ essential matrices $E_{a}$ are rectangular matrices of
dimension $s \times r$ defined by setting  $(E_{a})_{ib}  =  (F_{i})_{ab}$
(the reader should be cautious about the meaning of
indices: our indices $i$ or $a$ refer to actual vertices
of the graphs but the numbers chosen for labelling rows and columns
depend on some arbitrary ordering on these sets of vertices).
The particular matrix $E_{0}$ is usually called ``intertwiner'', in the
statistical physics literature;
it also describes ``essential paths'' emanating from the origin 
(we shall not need this notion in the present paper).
One can check that, for graphs with self-fusion,  $E_{a}=E_{0} G_{a}$.

Vertices of the diagram $G$ should be thought of as an
analogue of irreducible
representations for a subgroup of a group;  the
irreducible representations of the bigger group are themselves represented by
vertices of the graph $A$.
In this analogy, the first column of each matrix $F_{i}$ describes
the branching rule of $\tau_{i}$ with respect to the chosen subgroup
(restriction mechanism). In the same way, the columns of the
particular essential matrix $E_{0}$ describe the induction mechanism:
the non-zero matrix elements of the column labelled by $\sigma_{b}$
tell us what are those
representations $\tau_{i}$ that contain $\sigma_{b}$ in their
decomposition (for the branching $A \rightarrow G$).

Let
us recall how we compute the (twisted) partition functions $Z_{x,y}$,
at least, in the cases where $Oc(G) \simeq G\otimes_{J} G$. Again, we
follow the method explained in \cite{Coque:Qtetra} and
refer to \cite{CoqueGil:ADE} for a discussion of all the $ADE$
cases, but another formalism for calculating these quantities was
described in \cite{PetZub:Oc}.
The bimodule structure of $Oc(G)$ with respect to the corresponding 
${\cal A}_k$
algebra is encoded by matrices $W_{x,y}$ defined as
$\tau_i . x . \tau_j = \sum_y (W_{x,y})_{ij} . y$. One sets $W_x 
 =  W_{x,\ud{0}}$
and obtain the corresponding twisted partition functions as sesquilinear
forms $Z_{x,y} = \overline{\chi} W_{x,y} \chi$, or  $Z_x= Z_{x,\ud{0}}$.
Here $\chi$ is a vector in the complex vector space $\CC^s$.
The modular invariant partition function is $Z_{\ud{0}}$ with $\ud{0} 
= 0 \otimesdot 0$.
The $W_{x,y}$ can be simply obtained from the $W_{x}$ by working out the
multiplication table of $Oc(G)$ and decomposing the product $x \times
y$ on the basis generators (one of us (R.C.)
acknowledges discussions with M. Huerta about this).
Practically, once we have the $r$ rectangular
matrices $E_{a}$, of dimension $s \times r$ (with $s =\kappa-1$ for
$ADE$ diagrams), we first replace by $0$ all
the matrix elements of the columns labelled by vertices $b$ that {\sl
do not} belong to the subset $J$ of the graph $G$,
call $E_{a}^{red}$ these ``reduced'' matrices and obtain,
for each point $x = a\otimesdot b$ of the Ocneanu graph
$Oc(G)$ (in some cases, $x$ may be a linear combination
of such elements), a ``toric matrix''  $W_{x} = E_{a}\, (E_{b}^{red})^T$, of size
$s\times s$.

 The usual  partition function on a torus is calculated by identifying the states 
at the end of a cylinder through the trace operation. 
One may incorporate the action of an operator $X$ attached to a 
non -- trivial cycle of the cylinder before identifying the two ends.
This  operator should commute with
the Virasoro generators and its effect is basically to twist the
boundary conditions. An explicit expression, in the presence of two twists $X$ and $Y$,
was written for such a twisted partition function
  by \cite {PetZub:bcft},  \cite{PetZub:Oc}; it involves 
  matrix elements of the modular operator $S$. 
  Our own determination of the toric matrices (and corresponding twisted partition functions)  uses directly the fusion algebra -- \ie the graph algebra of the 
  $A_n$ diagrams.  Of course we could, by using the Verlinde formula, express the
  fusion rule coefficients through the matrix $S$, but in our approach, 
the diagrams themselves are taken as primary data and we do not need to use this operator at all, at least for the determination of the $W_{x,y}$.

\section {$ADE$ diagrams: the $su(2)$ system}

\subsection{Preliminary remarks}

$ADE$ Dynkin diagrams are well known. Their norm (highest eigenvalue of the
adjacency matrix) is
$2 \cos \left( \frac{ \pi}{\kappa} \right) = \beta$.
Diagrams $A_{\kappa -1}$ have $r$ points $\tau_{j}$, $j =
0,\ldots\kappa -2$, with $r=\kappa-1=k+1$ (this defines the level $k$).
In the light of McKay correspondance\cite{McKay}, these diagrams appear 
as quantum analogues of binary polyhedral groups \cite{Coque:Karpacz}, \cite{KiriOs},
\cite{Klein}.
For $su(2)$, the recurrence formula for adjacency matrices associated
with irreps is very well known: we have $N_0 N_j = N_j$, $N_1 N_j =
N_{j-1}+N_{j+1}$, for $1 \leq j \leq \kappa-2$.  This is the usual
multiplication of
spin $j/2$ representation by the fundamental (spin $1/2$).
For the diagram $A_{\kappa -1}$,
we also have a truncation of the spin rule: $N_1 N_{\kappa -2} =
N_{\kappa - 3}$.

Left action of the algebra $A_{\kappa - 1}$ on the vector space of a
diagram $G$ is defined by setting $F_{0}  =  G_{0} =
{\bf 1}_{r\times r}$,
     $F_{1}  =   G_{1}$, and compatiblility with left multiplication in
$A_{\kappa-1}$ is ensured by imposing the spin rule $F_{1} F_{i} =
F_{i-1} + F_{i+1}$,
a relation that determines the $F_{i}$'s iteratively.

The modular generator $T$, in the
Hurwitz-Verlinde representation, is given by
  $$T_{jj'} = \exp[2i\pi(\frac{(j+1)^2}{4 \kappa} -  \frac{1}{8})] 
\delta_{jj'} $$
where $j,j'$ run from $0$ to $\kappa -2$.

    The value of $T$ on the vertex $\tau_{j}$ of $A_{\kappa -1}$ is
    therefore determined, up to a global phase, by the quantity
    $\hat{T} =(j+1)^2$ mod $4 \kappa$, that we will call the ``modular
    exponent'' (see also the appendix).  The algebras of quantum
    symmetries $Oc(G)$, for diagrams $G$ of type ADE, are already
    known, and the corresponding Ocneanu graphs can be found for
    instance in \cite{Mercat-et-al}, \cite{Coque:Qtetra},
    \cite{CoqueGil:ADE}, \cite{Ocneanu:paths}, \cite{PetZub2},
    \cite{PetZub:Oc}, or also, in the context of the theory of
    induction of sectors, in \cite{Evans}.  In the present section,
    the overlap with \cite{CoqueGil:ADE} is important: in the later
    reference, an algebraic realization of the algebras $Oc(G)$ was
    given, but the primary data was the Ocneanu graph itself, taken
    from \cite{Ocneanu:paths}.  In the present section, our aim is
    neither to describe the algebras of quantum symmetries nor their
    corresponding graphs, since this is known already, but to show how
    the modular properties of the diagrams (in particular the table of
    eigenvalues for the operator $T$) together with the
    induction-restriction pattern, can be used to recover the known
    algebras of quantum symmetries.  This section also provides a kind
    of introduction to section 4 where the same techniques will be
    used to determine the structure of $Oc(G)$ for several diagrams
    belonging to the $su(3)$ system.

\subsection{First example: the $E_6$ case}

\begin{itemize}

\item Graphs.

The vector space of $E_6$ is both an associative (and commutative)
algebra with positive integral structure constants  (in other words,
it admits self-fusion), and it is
a module over $A_{11}$.
This example is fully studied in \cite{Coque:Qtetra} (see also \cite{Coque:Karpacz}); 
in particular its graph algebra matrices,  essential matrices,
Ocneanu graph and toric matrices are given there.
The $E_6$ Dynkin diagram  and the corresponding $A$ diagram with
same norm (\ie $A_{11}$) are displayed in Figure 1.

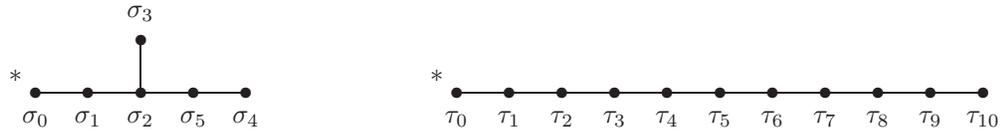
\begin{figure}[h]
\unitlength 0.7mm
\begin{center}
\begin{picture}(170,20)
\multiput(0,5)(10,0){5}{\circle*{2}}
\put(20,15){\circle*{2}}
\put(0,5){\line(1,0){40}}
\put(20,5){\line(0,1){10}}
\put(0,0){\makebox(0,0){$\sigma_0$}}
\put(10,0){\makebox(0,0){$\sigma_1$}}
\put(20,0){\makebox(0,0){$\sigma_2$}}
\put(30,0){\makebox(0,0){$\sigma_5$}}
\put(40,0){\makebox(0,0){$\sigma_4$}}
\put(20,20){\makebox(0,0){$\sigma_3$}}

\multiput(80,5)(10,0){11}{\circle*{2}}
\put(80,5){\line(1,0){100}}
\put(80,0){\makebox(0,0){$\tau_0$}}
\put(90,0){\makebox(0,0){$\tau_1$}}
\put(100,0){\makebox(0,0){$\tau_2$}}
\put(110,0){\makebox(0,0){$\tau_3$}}
\put(120,0){\makebox(0,0){$\tau_4$}}
\put(130,0){\makebox(0,0){$\tau_5$}}
\put(140,0){\makebox(0,0){$\tau_6$}}
\put(150,0){\makebox(0,0){$\tau_7$}}
\put(160,0){\makebox(0,0){$\tau_8$}}
\put(170,0){\makebox(0,0){$\tau_9$}}
\put(180,0){\makebox(0,0){$\tau_{10}$}}

\put(-5,7){$\ast$}
\put(75,7){$\ast$}

\end{picture}
\end{center}
\caption{The $E_6$ and $A_{11}$ Dynkin diagrams}
\end{figure}

For trees with one branching point (for instance
$E_6$, $E_7$ and $E_8$ diagrams), we label (one of) the longest
branches with increasing integers starting from $0$, up to
the branching point, then we jump to the extremity of the next
(clockwise) branch and so on. This is the ordering consistently
chosen in \cite{Coque:Qtetra} and \cite{CoqueGil:ADE}.

\item Restriction mechanism.

We look at $E_6$ as a module over $A_{11}$. For this, we
define an action of
$A_{11}$ on $E_6$:
$$
\begin{array}{ccl}
A_{11} \times E_6 &\rightarrow & E_6 \\
\tau_0 . \sigma_i &=& \sigma_i \\
\tau_1 . \sigma_i &=& \sum' \sigma_j
\end{array}
$$
where $\sum'$ runs over the neighbours of $\sigma_{i}$
on the diagram $E_{6}$. \\
We have obvious restrictions: $\tau_0 \hookrightarrow \sigma_0, \;
\tau_1 \hookrightarrow \sigma_1$. To obtain the others, we
impose the compatibility condition: $(\tau_1)^n . \sigma_i =
(\sigma_1)^n . \sigma_i$. We therefore calculate
the powers of the fundamentals $\tau_1$ and $\sigma_1$ and compare the results:
$$
\begin{array}{lll}
(\tau_1)^2 = \tau_0 + \tau_2, & (\sigma_1)^2 = \sigma_0 + \sigma_2, &
\textrm{so} \qquad  \tau_2 \hookrightarrow \sigma_2;\\
(\tau_1)^3 = 2\tau_1 + \tau_3, & (\sigma_1)^3 = 2\sigma_1 +
\sigma_3+\sigma_5, & \textrm{so} \qquad
\tau_3 \hookrightarrow \sigma_3 + \sigma_5; \\
\end{array}
$$
and so on. \\
In this way, we get the following branching rules 
$\tau_i \hookrightarrow \oplus E_{ij}^0 \sigma_j$ (essential matrix $E_0$):
$$
\begin{array}{llll}
\tau_0 \hookrightarrow \sigma_0 & \tau_1 \hookrightarrow \sigma_1 &
\tau_2 \hookrightarrow \sigma_2 & \tau_3
\hookrightarrow \sigma_3 + \sigma_5 \\
\tau_4 \hookrightarrow \sigma_2 + \sigma_4 & \tau_5 \hookrightarrow
\sigma_1 + \sigma_5 & \tau_6 \hookrightarrow \sigma_0 + \sigma_2
     & \tau_7 \hookrightarrow \sigma_1 + \sigma_3 \\
\tau_8 \hookrightarrow \sigma_2 & \tau_9 \hookrightarrow \sigma_5 &
\tau_{10} \hookrightarrow \sigma_4 & {}
\end{array}
$$

The rectangular $11 \times 6$ matrix $E_0$ encodes this result,
\ie the above branching rules give us the lines of this matrix.
Notice that this determination of $E_0$ does not require any
calculation involving essential paths (this
notion, although extremely nice and useful, is not required  at
this level).

Once the adjacency matrix $G_1$ is known (read it from the graph),
and the essential matrix $E_0$ (or intertwiner)
determined, we can use the general formulae given in the introduction
to determine the $6$ graph matrices $G_a$,
the $11$ matrices $F_i$ and the other essential matrices $E_a$ (six
of them, including $E_0$). Notice that, from the very beginning, we could
have proceeded
differently, determining first the $F_i$ by using both the equation $F_{1} =
G_{1}$ and the $su(2)$ rule of
composition of spins (recurrence relation); these matrices,
in turn, determine the $E_a$'s (in particular $E_0$).

\item Induction mechanism.

We now look at these previous branching rules, but in the opposite
direction: for instance $\sigma_3$ {\it comes from}
$\tau_3$ and $\tau_7$, so we can write $\sigma_3 \hookleftarrow
(\tau_3 , \tau_7)$. We get the induction correspondence
$E_6 \hookleftarrow A_{11}$ displayed in Fig \ref{E6/A11induction}.  
This is only another way to write the columns of the $E_0$ matrix.
We also plot  the values of the modular exponent $\hat T$ for the 
vertices $\tau_i$'s of $A_{11}$.

\begin{figure}[hhh]

\unitlength 0.7mm

\begin{center}
\begin{picture}(170,35)
\multiput(0,20)(10,0){5}{\circle*{2}}
\put(20,30){\circle*{2}}
\put(0,20){\line(1,0){40}}
\put(20,20){\line(0,1){10}}
\put(0,15){\makebox(0,0){$\tau_{0}$}}
\put(0,10){\makebox(0,0){$\tau_{6}$}}
\put(10,15){\makebox(0,0){$\tau_{1}$}}
\put(10,10){\makebox(0,0){$\tau_{5}$}}
\put(10,5){\makebox(0,0){$\tau_{7}$}}
\put(20,15){\makebox(0,0){$\tau_{2}$}}
\put(20,10){\makebox(0,0){$\tau_{4}$}}
\put(20,5){\makebox(0,0){$\tau_{6}$}}
\put(20,0){\makebox(0,0){$\tau_{8}$}}
\put(30,15){\makebox(0,0){$\tau_{3}$}}
\put(30,10){\makebox(0,0){$\tau_{5}$}}
\put(30,5){\makebox(0,0){$\tau_{9}$}}
\put(40,15){\makebox(0,0){$\tau_{4}$}}
\put(40,10){\makebox(0,0){$\tau_{10}$}}
\put(20,35){\makebox(0,0){$\tau_{3},\tau_{7}$}}

\put(0,20){\circle{4}}
\put(20,30){\circle{4}}
\put(40,20){\circle{4}}

\multiput(80,20)(10,0){11}{\circle*{2}}
\put(80,20){\line(1,0){100}}
\put(80,15){\makebox(0,0){$\tau_0$}}
\put(90,15){\makebox(0,0){$\tau_1$}}
\put(100,15){\makebox(0,0){$\tau_2$}}
\put(110,15){\makebox(0,0){$\tau_3$}}
\put(120,15){\makebox(0,0){$\tau_4$}}
\put(130,15){\makebox(0,0){$\tau_5$}}
\put(140,15){\makebox(0,0){$\tau_6$}}
\put(150,15){\makebox(0,0){$\tau_7$}}
\put(160,15){\makebox(0,0){$\tau_8$}}
\put(170,15){\makebox(0,0){$\tau_9$}}
\put(180,15){\makebox(0,0){$\tau_{10}$}}

\put(70,6){\makebox(0,0){$\hat{T}:$}}

\put(80,5){\makebox(0,0){ 1}}
\put(90,5){\makebox(0,0){4}}
\put(100,5){\makebox(0,0){9}}
\put(110,5){\makebox(0,0){ 16}}
\put(120,5){\makebox(0,0){25}}
\put(130,5){\makebox(0,0){36}}
\put(140,5){\makebox(0,0){ 1}}
\put(150,5){\makebox(0,0){ 16}}
\put(160,5){\makebox(0,0){33}}
\put(170,5){\makebox(0,0){4}}
\put(180,5){\makebox(0,0){ 25}}

\end{picture}
\end{center}
\caption{The $E_6 \hookleftarrow A_{11}$ induction graph and the values
of $\hat T$ on irreps of $A_{11}$}
\label{E6/A11induction}
\end{figure}
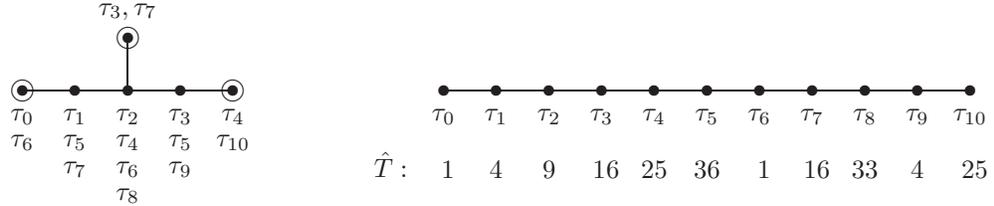

From the induction graph we have: $\sigma_{0} \hookleftarrow
(\tau_{0}, \tau_{6})$, and
we notice that the value of the modular matrix $T$ on $\tau_{0}$ and
$\tau_{6}$ is the same
(also for  $\tau_{3}$ and $\tau_{7}$, and for  $\tau_{4}$ and $\tau_{10}$).
This allows one to assign a fixed value of $T$ to three particular
vertices of $E_6$:
$\hat{T}(\sigma_{0})=1$, $\hat{T}(\sigma_3) = 16$ and $\hat{T}(\sigma_4) = 25$.
For every other point of the $E_6$ graph, the value of $T$ that would
be inherited from the $A_{11}$ graph
by this induction mechanism is not uniquely determined  (for
instance, in the case of $\sigma_1$, the
values of $\hat{T}$ obtained from $A_{11}$ would be  associated with
$\tau_1, \tau_5$ and $\tau_7$
but these values are not all equal).
These elements $\{\sigma_{0}, \sigma_{4}, \sigma_{3} \}$ span the
subalgebra $J=A_3$.
This subalgebra is known to admit an invariant supplement in the
graph algebra of $E_6$.

\item Quantum symmetries.

The Ocneanu graph of $E_6$ given in \cite{Coque:Qtetra}, \cite{CoqueGil:ADE}, 
\cite{Ocneanu:paths}, \cite{PetZub:Oc}, is the
Cayley graph of multiplication
by the two generators of an associative algebra $Oc(E_6)$ which can be
realized (see \cite{Coque:Qtetra}
and \cite{CoqueGil:ADE}) as $E_6 \otimes_{A_3} E_6$. It has $12=6 \times 6 /3$ vertices,
three of them being ambichiral, namely $\sigma_0 \otimesdot \sigma_0$, $\sigma_0 \otimesdot \sigma_3$
and $\sigma_0 \otimesdot \sigma_4$.
We introduce
the symbol $\otimesdot$ to denote $\otimes_{A_3}$
to stress the fact that
the tensor product is taken not above the complex numbers but above
the subalgebra $J=A_3$.
This means that $a \, u \otimesdot b = a \otimesdot u\, b$ whenever
$u \in \{\sigma_{0}, \sigma_{4}, \sigma_{3} \}$
and $a, b \in E_6$.
The point that we make, here, is that this subalgebra $A_3$ is
actually {\sl determined} as above,
by induction, from the eigenvalues of the $T$ operator.

\item Dimensions of blocks

Diagonalization of the two algebra structures of ${\cal B}{E_6}$ leads
to the quadratic sum rule 
$$
dim({\cal B}{E}_6) = \sum_{i\in A_{11}} d_i^2 = \sum_{x \in Oc(E_6)} d_x^2 
 = 2512 = (2)^4 (157)^1
$$
where $d_i = \sum_{a,b \in G} (F_{i})_{ab}$ runs in the list
$(6,10,14,18,20,20,20,18,14,10,6)$ and where, for $x = a \otimesdot
b$, $d_x = \sum_{i,j \in A_{11}} (G_a.G_b)_{ij}$ runs in the list
$(6,8,6,10,14,10,10,14,10,20,28,20)$.  This identity follows directly
from the fact that ${\cal B}{E_6}$ can be written in two different
ways as a direct sum of matrix algebras (${\cal B}{E_6}$ is
semi-simple for both structures).  \\
We have also the linear sum rule $\sum d_i = \sum d_x = 720 = (2)^4
(3)^2 (5)^1$.  Such a linear sum rule also holds ``experimentally'' in
almost all $ADE$ cases (for the $D_{2n}$ cases one has actually to
introduce a simple correcting factor, as explained in
\cite{PetZub:Oc}).  In general, there is no reason, for a general
bialgebra --even semi-simple for both structures-- to give rise to such
a linear sum rule.  The interpretation of this property is therefore
still mysterious.  As we shall see in the next part, it
also holds for the several examples of diagrams of type $su(3)$ that we
have analysed so far.

There are also quantum sum rules (``mass relations''): define $o(G)
\doteq \sum_{a \in G} qdim_{a}^{2}$, where $qdim_{a}$ are the quantum
dimensions of the vertices $a$ of $G$ (for example $o(E_{6})=4(3+\sqrt
3)$, $o(A_{11})=24 (2 + \sqrt 3)$, $o(A_{3}) = 1 + (\sqrt 2))^{2} =
4)$; then, if the diagonalizations of the two algebra structures of
${\cal B}(G)$ are described respectively by ${\cal A}_{k}$, for some 
$k$, and by $Oc(G) = G \otimes_{J} G$ for some $J$, one can check that
$o(Oc(G))$ defined as $\frac{o(G) \times
o(G)}{o(J)}$ is equal to $o({\cal A}_{k})$. In the present case,  $\frac{o(E_{6})
\times o(E_{6})}{ o(A_{3})} = o(A_{11})$.
This observational fact, properly generalized,  holds for all ADE 
diagrams.
Indeed,  $ o(D_n)   = \frac{1}{2} o(A_{2n-3})$ and  $\frac{o(E8) 
\times o(E8)}{o(J)} = o(A_{29})$,
where $o(J) = [1]_q^2 +\left(\frac{[5]_q}{[3]_q}\right)^2$  since the 
quantum dimensions of  vertices
$\sigma_0$ and $\sigma_6$ spanning the subspace $J$ of $E_8$ are respectively  
equal to the q-numbers
$[1]_q$ and $\frac{[5]_q}{[3]_q}$ (here $q = \exp(\frac{i \pi}{30})$). 
In the case of $E_7$, we found that $o(Oc(E_7))$ defined as $\frac{o(D_{10})
\times o(D{10})}{o(J)}$, where $J$ is the subalgebra of $D_{10}$, is equal 
to $o(A_{17})$.
We found also empirically  the relation
$o(A_{17}) = \frac{o(E_7) \times o(D_{10})}{ o(J)}$, where $o(J) = [2]_q^2 +  [4]_q^2 + 
\left( \frac{[4]_q}{[3]_q}\right)^2$ and where
the q-numbers $[2]_q$,  $[4]_q$ and $\frac{[4]_q}{[3]_q}$ are the q-dimensions of the 
vertices $\sigma_1$, $\sigma_3$ and $\sigma_5$ of $E_7$ (here $q = \exp(\frac{i \pi}{18})$). 
Analoguous quantum sum rules hold for
the several examples of diagrams of type $su(3)$ that we have analysed so 
far. We do not know any general formal proof of these quantum relations.

\end{itemize}

%%%%%%%%%%%%%%%%%%%%%%%%%%

\subsection{The $ADE$ diagrams}

We show in this section how all cases relative to the $su(2)$ system can be studied in the same manner.

\begin{itemize}

\item $E_6$ case.\\ It was studied in the last section.

\item $E_8$ case.\\ The cases of $E_6$ and $E_8$ are very similar.
The Dynkin diagram of the $A$ series with same Coxeter number
($\kappa = 30$) as $E_8$ is $A_{29}$.
Like $E_6$, the vector space of the diagram $E_8$ admits self-fusion
(associative algebra structure with positive integral structure constants).

\begin{figure}[hhh]
\unitlength 0.7mm
\begin{center}

\begin{picture}(90,80)(0,3)
\thinlines
\multiput(0,55)(15,0){7}{\circle*{2}}
\put(60,70){\circle*{2}}
\put(0,55){\circle{4}}
\put(90,55){\circle{4}}
\put(-8,55){$\ast$}
\thicklines
\put(0,55){\line(1,0){90}}
\put(60,55){\line(0,1){15}}

\put(0,60){\makebox(0,0){$\bf{\sigma_{0}}$}}
\put(15,60){\makebox(0,0){$\bf{\sigma_{1}}$}}
\put(30,60){\makebox(0,0){$\bf{\sigma_{2}}$}}
\put(45,60){\makebox(0,0){$\bf{\sigma_{3}}$}}
\put(64,60){\makebox(0,0){$\bf{\sigma_{4}}$}}
\put(75,60){\makebox(0,0){$\bf{\sigma_{7}}$}}
\put(90,60){\makebox(0,0){$\bf{\sigma_{6}}$}}
\put(60,80){\makebox(0,0){$\bf{\sigma_{5}}$}}

\put(0,50){\makebox(0,0){$\tau_{0}$}}
\put(0,45){\makebox(0,0){$\tau_{10}$}}
\put(0,40){\makebox(0,0){$\tau_{18}$}}
\put(0,35){\makebox(0,0){$\tau_{28}$}}

\put(15,50){\makebox(0,0){$\tau_{1}$}}
\put(15,45){\makebox(0,0){$\tau_{9}$}}
\put(15,40){\makebox(0,0){$\tau_{11}$}}
\put(15,35){\makebox(0,0){$\tau_{17}$}}
\put(15,30){\makebox(0,0){$\tau_{19}$}}
\put(15,25){\makebox(0,0){$\tau_{27}$}}

\put(30,50){\makebox(0,0){$\tau_{2}$}}
\put(30,45){\makebox(0,0){$\tau_{8}$}}
\put(30,40){\makebox(0,0){$\tau_{10}$}}
\put(30,35){\makebox(0,0){$\tau_{12}$}}
\put(30,30){\makebox(0,0){$\tau_{16}$}}
\put(30,25){\makebox(0,0){$\tau_{18}$}}
\put(30,20){\makebox(0,0){$\tau_{20}$}}
\put(30,15){\makebox(0,0){$\tau_{26}$}}

\put(45,50){\makebox(0,0){$\tau_{3}$}}
\put(45,45){\makebox(0,0){$\tau_{7}$}}
\put(45,40){\makebox(0,0){$\tau_{9}$}}
\put(45,35){\makebox(0,0){$\tau_{11}$}}
\put(45,30){\makebox(0,0){$\tau_{13}$}}
\put(45,25){\makebox(0,0){$\tau_{15}$}}
\put(45,20){\makebox(0,0){$\tau_{17}$}}
\put(45,15){\makebox(0,0){$\tau_{19}$}}
\put(45,10){\makebox(0,0){$\tau_{21}$}}
\put(45,5){\makebox(0,0){$\tau_{25}$}}

\put(60,50){\makebox(0,0){$\tau_{4}$}}
\put(60,45){\makebox(0,0){$\tau_{6}$}}
\put(60,40){\makebox(0,0){$\tau_{8}$}}
\put(60,35){\makebox(0,0){$\tau_{10}$}}
\put(60,30){\makebox(0,0){$\tau_{12}$}}
\put(60,25){\makebox(0,0){$2\tau_{14}$}}
\put(60,20){\makebox(0,0){$\tau_{16}$}}
\put(60,15){\makebox(0,0){$\tau_{18}$}}
\put(60,10){\makebox(0,0){$\tau_{20}$}}
\put(60,5){\makebox(0,0){$\tau_{22}$}}
\put(60,0){\makebox(0,0){$\tau_{24}$}}

\put(75,50){\makebox(0,0){$\tau_{5}$}}
\put(75,45){\makebox(0,0){$\tau_{7}$}}
\put(75,40){\makebox(0,0){$\tau_{11}$}}
\put(75,35){\makebox(0,0){$\tau_{13}$}}
\put(75,30){\makebox(0,0){$\tau_{15}$}}
\put(75,25){\makebox(0,0){$\tau_{17}$}}
\put(75,20){\makebox(0,0){$\tau_{21}$}}
\put(75,15){\makebox(0,0){$\tau_{23}$}}

\put(90,50){\makebox(0,0){$\tau_{6}$}}
\put(90,45){\makebox(0,0){$\tau_{12}$}}
\put(90,40){\makebox(0,0){$\tau_{16}$}}
\put(90,35){\makebox(0,0){$\tau_{22}$}}

\put(60,75){\makebox(0,0){$\tau_{5},\tau_{9},\tau_{13},\tau_{15},\tau_{19},
\tau_{23}$}}

\end{picture}

\end{center}
\caption{The $E_8 \hookleftarrow A_{29}$ induction graph }
\end{figure}
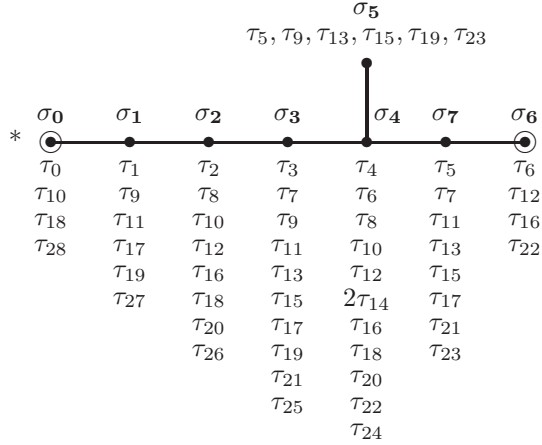

The value of $\hat{T}$ on irreps
$(\tau_0,\tau_1,\tau_2, \cdots,\tau_{28})$ of $A_{29}$
(equal for $\tau_j$ to $(j+1)^2$ mod 120) gives:
$$
(\ud{1},4,9,16,25,36,\udd{49},64,81,100, 
\ud{1},22,\udd{49},76,105,16,\udd{49},84,\ud{1},40,
81,4,\udd{49},96,25,76,9,64,\ud{1})
$$
We see that $T$ has the same value on vertices $\tau_{j}$ that
correspond to $\sigma_{0}$ ($\hat{T}=1$). Same comment for $\sigma_{6}$ ($\hat{T}=49$).
We therefore take $J = \{\sigma_{0}, \sigma_{6}\}$; this generates a
subalgebra which is isomorphic with the algebra of the $A_{2}$ graph.
We have indeed $Oc(E_{8}) = E_{8}\otimes_{A_{2}} E_{8}$ and the
Ocneanu graph has $32 = 8\times 8/2$ vertices, two of them being ambichiral,
namely $\sigma_{0} \otimesdot \sigma_{0}$ and $\sigma_{0} \otimesdot
\sigma_{6}$. Notice that $\sigma_{6} \otimesdot \sigma_{6} = \sigma_{0}
\otimesdot \sigma_{0}.$ 
Dimensions of blocks can be computed as before (see for instance \cite{CoqueGil:ADE}).
One writes $dim ({\cal B}E_8) = 63136 = (2)^5 (1973)^1$ in two different ways as a sum of $29$
or $32$ squares.
The linear sum rule gives $\sum d_i = \sum d_x = 1240 = (2)^3 (5)^1 (31)^1$.

\item $A_{\kappa -1}$ cases\\ 
The induction-restriction rules from  $A_{\kappa -1}$ to itself are
of course trivial and the subalgebra $J$
determined by the constancy of $T$ on pre-images is equal
to the algebra  $A_{\kappa -1}$ itself.
The algebra $Oc(A_{\kappa -1})$ equal to  $A_{\kappa -1} \otimes_{A_{\kappa -1}}
A_{\kappa -1}$ is therefore isomorphic with  $A_{\kappa -1}$
itself. The Ocneanu graph coincides with the original Dynkin diagram.

\item $D_{2n+1}$ cases\\ 
The Dynkin diagram of the $A$ series with same Coxeter number ($\kappa = 4n-2$)
as  $D_{2n+1}$ is $A_{4n-1}$.
Actually (see \cite{orbifolds}), $D$ diagrams are $\ZZ_2$ orbifolds of $A$
diagrams.

Let's first have a look at the $A_7$ case. Its Dynkin diagram and the values of
$\hat{T}$ on irreps $\tau_i$'s are given in Fig \ref{A7diagram}.
\begin{figure}[hhh]
\unitlength 0.7mm
\begin{center}
\begin{picture}(90,10)(0,5)
\put(0,15){\line(1,0){90}}
\multiput(0,15)(15,0){7}{\circle*{2}}
\put(0,7.5){\makebox(0,0){$\tau_{0}$}}
\put(15,7.5){\makebox(0,0){$\tau_{1}$}}
\put(30,7.5){\makebox(0,0){$\tau_{2}$}}
\put(45,7.5){\makebox(0,0){$\tau_{3}$}}
\put(60,7.5){\makebox(0,0){$\tau_{4}$}}
\put(75,7.5){\makebox(0,0){$\tau_{5}$}}
\put(90,7.5){\makebox(0,0){$\tau_{6}$}}
\put(-10,1){\makebox(0,0){$\hat{T}:$}}

\put(0,0){\makebox(0,0){1}}
\put(15,0){\makebox(0,0){4}}
\put(30,0){\makebox(0,0){9}}
\put(45,0){\makebox(0,0){16}}
\put(60,0){\makebox(0,0){25}}
\put(75,0){\makebox(0,0){4}}
\put(90,0){\makebox(0,0){17}}

\end{picture}
\end{center}
\caption{The $A_7$ diagram and the values of $\hat T$}
\label{A7diagram}
\end{figure}
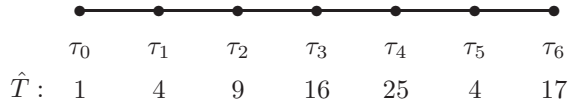

The algebra of quantum symmetries of $A_7$, as we saw, is $A_7
\otimes_{A_7} A_7
\simeq A_7$,  but there is also another
way to quotient the tensor product if we want $T$ to be well defined in the
quotient.
We see that the values of $T$ are the same ($\hat T = 4$) for $\tau_1$
and $\tau_5$, so  we define therefore a map  (twist)
$\rho: A_7 \rightarrow A_7$ such that:
$$
\rho(\tau_i) = \tau_i \qquad \text{for} \;\, i \in \{0,2,3,4,6 \}
\qquad \text{and} \quad
\rho(\tau_1) = \tau_5, \quad \rho(\tau_5) = \tau_1.
$$
Defining then $Oc(D_5) = A_7 \otimes_{\rho(A_7)} A_7$ we
recover the algebra of quantum symmetry of $D_5$.\\
This can be generalized for all $D_{2n+1}$ cases. These diagrams do not enjoy self-fusion.\\

\item $D_{2n}$ cases\\
Starting with the $D_{2n}$ diagram and graph algebra, we obtain the
following induction-restriction graph with respect to
the corresponding $A$ diagram with the same norm ($A_{4n-3}$).

\begin{figure}[hhh]
\unitlength 0.9mm
\begin{center}
\begin{picture}(130,25)
\put(0,10){\line(1,0){40}}
\put(70,10){\line(1,0){40}}
\put(110,10){\line(3,2){15}}
\put(110,10){\line(3,-2){15}}
\multiput(0,10)(20,0){3}{\circle*{2}}
\multiput(70,10)(20,0){3}{\circle*{2}}
\put(125,20){\circle*{2}}
\put(125,0){\circle*{2}}

\put(0,10){\circle{3.5}}
\put(40,10){\circle{3.5}}
\put(90,10){\circle{3.5}}
\put(125,20){\circle{3.5}}
\put(125,0){\circle{3.5}}

\dashline[30]{2}(40,10)(70,10)

\small
\put(0,5){\makebox(0,0){$\tau_{0}$}}
\put(0,0){\makebox(0,0){$\tau_{4n-4}$}}
\put(0,15){\makebox(0,0){$\sigma_{0}$}}
\put(20,5){\makebox(0,0){$\tau_{1}$}}
\put(20,0){\makebox(0,0){$\tau_{4n-5}$}}
\put(20,15){\makebox(0,0){$\sigma_{1}$}}
\put(40,5){\makebox(0,0){$\tau_{2}$}}
\put(40,0){\makebox(0,0){$\tau_{4n-6}$}}
\put(40,15){\makebox(0,0){$\sigma_{2}$}}
\put(70,5){\makebox(0,0){$\tau_{2n-5}$}}
\put(70,0){\makebox(0,0){$\tau_{2n+1}$}}
\put(70,15){\makebox(0,0){$\sigma_{2n-5}$}}
\put(90,5){\makebox(0,0){$\tau_{2n-4}$}}
\put(90,0){\makebox(0,0){$\tau_{2n}$}}
\put(90,15){\makebox(0,0){$\sigma_{2n-4}$}}
\put(108,5){\makebox(0,0){$\tau_{2n-3}$}}
\put(108,0){\makebox(0,0){$\tau_{2n-1}$}}
\put(108,15){\makebox(0,0){$\sigma_{2n-3}$}}
\put(132,23){\makebox(0,0){$\sigma_{2n-2}$}}
\put(132,17){\makebox(0,0){$\tau_{2n-2}$}}
\put(132,5){\makebox(0,0){$\sigma_{2n-2}^{'}$}}
\put(132,-3){\makebox(0,0){$\tau_{2n-2}$}}

\normalsize

\end{picture}
\end{center}
\caption{The $D_{2n}$-$A_{4n-3}$ induction graph}
\label{D2n/A4n-3induction}
\end{figure}
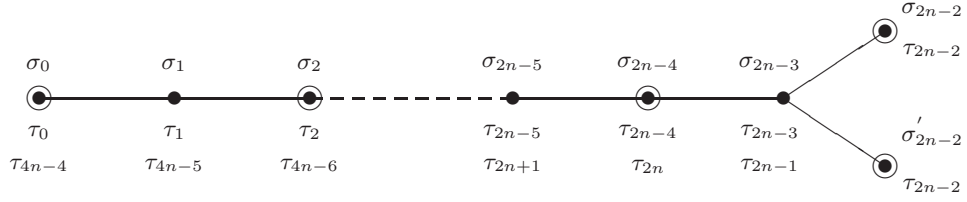

The value of $\hat T$ on the irreps $\{ \tau_0, \tau_1, \tau_2,
\cdots, \tau_{2n-1}, \tau_{2n}, \tau_{2n+1}, \cdots, \tau_{4n-6},
\tau_{4n-5}, \tau_{4n-4} \}$ of $A_{4n-3}$ gives:
$$
T(\tau_0) = T(\tau_{4n-4}) \qquad
T(\tau_2) = T(\tau_{4n-6}) \qquad \cdots \qquad
T(\tau_{2n-4}) = T(\tau_{2n})
$$
These last values are symmetric with respect to the central vertex $\tau_{2n}$.\\
We can assign a fixed value of $T$ for the irreps $\{ \sigma_0,
\sigma_2, \cdots, \sigma_{2n-4}, \sigma_{2n-2},
\sigma_{2n-2}^{'} \}$ for $D_{2n}$ (marked with a circle in the induction diagram).
They span the subalgebra $J$. However, we notice immediately that
something special happens here: the two ends of the fork (vertices
$\sigma_{2n-2}$ and $\sigma_{2n-2}^{'}$) are not distinguished by the
values of $T$. Actually, the determination of the graph matrices
$G_{a}$ for the Dynkin diagram $D_{2n}$ is not as straightforward as
for some other cases: looking for an associative algebra determined
by this diagram leads to a two-parameter family of solutions, but
there is only one solution (up to permutation $G_{2n-2} \leftrightarrow
G_{2n-2}^{'}$) that has correct self fusion, \ie integrality and
positivity of structure constants (a similar phenomenon
appears, for example, for the ${\cal E}_{9}$ diagram of the $su(3)$ system).
Since $T$ may be defined on any linear combination of these two
vertices, it is natural to expect that this arbitrariness is encoded,
at the level of the algebra of quantum symmetries, in a ``non-commutative 
geometrical spirit'', by an algebra of
$2\times 2$ matrices. $Oc(D_{2n})$ consists indeed of two separate
components: the first (usual) is given by $D_{2n}^{trunc}\otimes_{J^{'}}
D_{2n}^{trunc}$, where $D_{2n}^{trunc}$ is the vector space
corresponding to the subdiagram spanned by $\{\sigma_{0}, \sigma_{1}, \sigma_{2},
\ldots \sigma_{2n-3} \}$, obtained by removing the fork, and $J' =
\{\sigma_{0}, \sigma_{2},
\ldots \sigma_{2n-4} \}$ is the corresponding truncated subset of $J$.
The second component is a non-commutative $2\times 2$ matrix algebra
reflecting the indistinguishability of $\sigma_{2n-2}$ and $\sigma_{2n-2}^{'}$.
Ambichiral points are associated with the $n+1$ vertices of $J$ (\ie
$n-1$ for the linear branch and $2$ for the fork);
   we  expect therefore that the Ocneanu graph of $D_{2n}$ will have
$\frac{(2n-2)(2n-2)}{n-1} + 4 = 4n$ vertices. We could as well say
that the number of ``effective'' points of $J$ is $n$, rather than
$n+1$ and notice that $4n = \frac{2n \times 2n}{n}$. This is indeed
correct (see \cite{CoqueGil:ADE}, \cite{Ocneanu:paths}, \cite{PetZub:Oc}).
One way to realize the algebra $Oc(D_{2n})$ is to write it as a quotient of the 
semi-direct product, by $\ZZ_2 = \{-,+\}$ of the tensor 
square of the graph algebra $D_{2n}$.  
The non-commutativity of the multiplication can be seen, for instance,
 from the fact that $(2\otimesdot 0, +)(0 \otimesdot 0,-)= (2 \otimesdot 0, 
-)$, but
 $(0\otimesdot 0, -)(2 \otimesdot 0,+)= (2' \otimesdot 0, -)$.
The reader may refer to \cite{CoqueGil:ADE} for another explicit realization
of this algebra. In any case, the method followed so far, which is based on the 
eigenvalues of the $T$ operator, seems to be
insufficient to fully determine the Ocneanu graph in that example.

\item $E_7$ case (related to the $D_{10}$ case)\\
For the $D_{10}$ case, something special happens. The corresponding
$A$ diagram with the same
norm is $A_{17}$, and the value of $T$ in irreps $\{ \tau_0, \tau_1,
\cdots, \tau_8, \cdots, \tau_{15}, \tau_{16} \}$ of $A_{17}$ are:
$$
(1,4,\ud{9},16,25,36,49,64,\ud{9},28,49,0,25,52,\ud{9},40,1)
$$

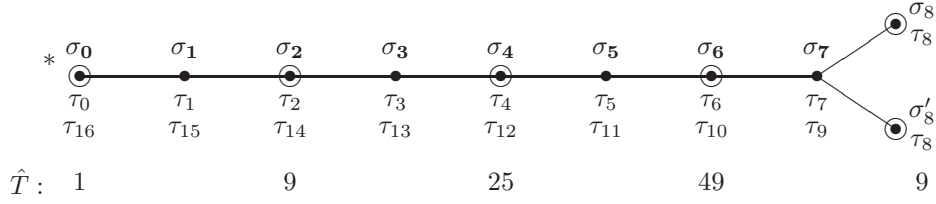
\begin{figure}[hhh]
\unitlength 0.7mm
\begin{center}
\begin{picture}(160,25)(0,-5)
\put(0,15){\line(1,0){140}}
\put(140,15){\line(3,2){15}}
\put(140,15){\line(3,-2){15}}
\multiput(0,15)(20,0){8}{\circle*{2}}
\put(155,25){\circle*{2}}
\put(155,5){\circle*{2}}
\put(155,25){\circle{4}}
\put(155,5){\circle{4}}
\multiput(0,15)(40,0){4}{\circle{4}}

\put(0,10){\makebox(0,0){$\tau_{0}$}}
\put(0,5){\makebox(0,0){$\tau_{16}$}}
\put(20,10){\makebox(0,0){$\tau_{1}$}}
\put(20,5){\makebox(0,0){$\tau_{15}$}}
\put(40,10){\makebox(0,0){$\tau_{2}$}}
\put(40,5){\makebox(0,0){$\tau_{14}$}}
\put(60,10){\makebox(0,0){$\tau_{3}$}}
\put(60,5){\makebox(0,0){$\tau_{13}$}}
\put(80,10){\makebox(0,0){$\tau_{4}$}}
\put(80,5){\makebox(0,0){$\tau_{12}$}}
\put(100,10){\makebox(0,0){$\tau_{5}$}}
\put(100,5){\makebox(0,0){$\tau_{11}$}}
\put(120,10){\makebox(0,0){$\tau_{6}$}}
\put(120,5){\makebox(0,0){$\tau_{10}$}}
\put(140,10){\makebox(0,0){$\tau_{7}$}}
\put(140,5){\makebox(0,0){$\tau_{9}$}}

\put(0,20){\makebox(0,0){$\bf{\sigma_{0}}$}}
\put(20,20){\makebox(0,0){$\bf{\sigma_{1}}$}}
\put(40,20){\makebox(0,0){$\bf{\sigma_{2}}$}}
\put(60,20){\makebox(0,0){$\bf{\sigma_{3}}$}}
\put(80,20){\makebox(0,0){$\bf{\sigma_{4}}$}}
\put(100,20){\makebox(0,0){$\bf{\sigma_{5}}$}}
\put(120,20){\makebox(0,0){$\bf{\sigma_{6}}$}}
\put(140,20){\makebox(0,0){$\bf{\sigma_{7}}$}}

\put(160,27.5){\makebox(0,0){$\sigma_{8}$}}
\put(160,22.5){\makebox(0,0){$\tau_{8}$}}

\put(160,8.5){\makebox(0,0){$\sigma_{8}'$}}
\put(160,2.5){\makebox(0,0){$\tau_{8}$}}

\put(-10,-5){\makebox(0,0){$\hat{T}:$}}
\put(0,-5){\makebox(0,0){$1$}}
\put(40,-5){\makebox(0,0){$9$}}
\put(80,-5){\makebox(0,0){$25$}}
\put(120,-5){\makebox(0,0){$49$}}
\put(160,-5){\makebox(0,0){$9$}}

\put(-7,17){$\ast$}

\end{picture}
\end{center}
\caption{The $D_{10}$-$A_{17}$ induction graph and the values of $\hat{T}$}
\end{figure}

These values are symmetric with respect to the central vertex, as in all
$A_{4n-3}$ case. For $A_{17}$, 
the value of $T$ on the central vertex ($\tau_8$) is equal to the value of $T$
on other vertices, namely
$\tau_2$ and $\tau_{14}$. 
This gives us another way to define a twist $\rho$ 
acting on the vertices of $D_{10}$ (this is ``the''
exceptional twist of the $su(2)$ Coxeter-Dynkin system; existence of this 
twist is not new, but what we discuss here is its relation
with  the modular  $T$ operator). In other words, we form the
tensor product $D_{10} \otimes D_{10}$, but
identify $a u \otimes b$ with $a \otimes \rho(u) b$ when $u\in
\{\sigma_0, \sigma_2, \sigma_4, \sigma_6, \sigma_8, \sigma_8'\}$
and
$$
\begin{array}{ccc}
\rho(\sigma_0)=\sigma_0, & \qquad \qquad \rho(\sigma_4)=\sigma_4, &
\qquad \qquad \rho(\sigma_8)=\sigma_2,  \\
\rho(\sigma_2)=\sigma_8, & \qquad \qquad \rho(\sigma_6)=\sigma_6, &
\qquad \qquad \rho(\sigma_8^{'})=\sigma_8^{'}.
\end{array}
$$

We obtain the algebra $D_{10} \otimes_{\rho} D_{10}$ which is
isomorphic with the algebra of quantum symmetries of the $E_7$
diagram.  The diagram $E_7$ does not enjoy self-fusion.

\end{itemize}

{\bf \ud{Remark $1$}}: The reader will have noticed that
we do not necessarily {\sl start} from a given
graph $G$ (for instance $E_7$), for which we want to deduce $Oc(G)$.
Rather, we first consider all those graphs $G$ which admit a good
algebra structure (self fusion), \ie $A$, $D_{2n}$, $E_6$ and
$E_8$; we then determine, for every one of them,
the induced pattern of $T$ eigenvalues by looking at the well
determined $A \rightarrow G$ restriction; finally,
we build all the possible quotients of $G \otimes G$ 
over the subalgebras -- and possibly twists --  determined by the pattern
of $T$ values.
For example, if we assume that $E_7$ is already known to ``exist''
(as a module over $A_{17}$), and since it does not admit self-fusion, the \
only thing that we expect a priori is
that its algebra of quantum symmetries $Oc(E_7)$ will be obtained as
a quotient of a tensor product of the
algebras $A_{17}$ or $D_{10}$. Therefore,
$Oc(E_7)$ is only the {\sl name} given to  $D_{10} \otimes_{\rho} D_{10}$; 
the graph $E_7$ itself can then be recognized as one of the two  subsets of vertices
of $Oc(E_7)$ that linearly generates a module over one of the two chiral parts
of the Ocneanu graph (each one being isomorphic with the algebra of $D_{10}$).

{\bf \ud{Remark $2$}}: As discussed previously, the method that we follow seems
to be insufficient to fully determine $Oc(G)$ when the later is not
commutative (cases when a coefficient strictly larger than $1$ appears
in the corresponding expression of the modular invariant partition
function). These is only one example of this kind for the $su(2)$ system (the
$D_{2n}$ diagrams), but there are several such examples for the $su(3)$ system.

%%%%%%%%%%%%%%%%%%%%%%%%%%%

\section {Di Francesco - Zuber diagrams: the $su(3)$ system}

\subsection {Preliminary remarks}

In the $su(2)$ case, the classification follows an $ADE$ pattern.
For $su(n), n\geq 2$ cases, there was no at-hand diagrams
to start with, but the
list of $su(3)$ diagrams (``generalized Coxeter Dynkin diagrams'') was
obtained in 1989 (with CAF $=$ Computer-Aided Flair)
by Di Francesco and  Zuber in \cite{DiFZub}; this list was
later shown to be complete by A. Ocneanu,
during the Bariloche school at the very beginning of 2000 (actually one of their graphs -- the one
called ${\cal E}_{3}^{(12)}$ in \cite{DiFZub} -- had to be removed).

Pictures of the graphs belonging to the  Coxeter-Dynkin system of
$su(3)$ can be found in
\cite{DiFZub},\cite{Zub:integrable},\cite{Zub:rootsystems},\cite{Zub:reflections}, and
in the book \cite{CFTbook}; we refer to  \cite{Ocneanu:Bariloche} and \cite{Zub:Bariloche}
or to the school web page
$www.univ-mrs.fr/\sim coque/Bariloche.html$ for the final list.
We do not discuss the $su(4)$ system in this paper, but these 
graphs can also be found in the
Ocneanu contribution to the same Bariloche school
\cite{Ocneanu:Bariloche} and on the corresponding web pages.

\begin{figure}[hhh]

%\unitlength 0.25mm
\begin{center}
\begin{picture}(210,160)
\put(0,0){\begin{picture}(40,40)
\put(0,0){\circle*{4}}
\put(40,0){\circle*{4}}
\put(20,30){\circle*{4}}
\put(0,0){\vector(1,0){22.5}}
\put(20,0){\line(1,0){20}}
\put(40,0){\vector(-2,3){11.5}}
\put(30,15){\line(-2,3){10}}
\put(20,30){\vector(-2,-3){11.5}}
\put(10,15){\line(-2,-3){10}}\end{picture}}
\put(20,30){\begin{picture}(40,40)
\put(0,0){\circle*{4}}
\put(40,0){\circle*{4}}
\put(20,30){\circle*{4}}
\put(0,0){\vector(1,0){22.5}}
\put(20,0){\line(1,0){20}}
\put(40,0){\vector(-2,3){11.5}}
\put(30,15){\line(-2,3){10}}
\put(20,30){\vector(-2,-3){11.5}}
\put(10,15){\line(-2,-3){10}}\end{picture}}
\put(40,0){\begin{picture}(40,40)
\put(0,0){\circle*{4}}
\put(40,0){\circle*{4}}
\put(20,30){\circle*{4}}
\put(0,0){\vector(1,0){22.5}}
\put(20,0){\line(1,0){20}}
\put(40,0){\vector(-2,3){11.5}}
\put(30,15){\line(-2,3){10}}
\put(20,30){\vector(-2,-3){11.5}}
\put(10,15){\line(-2,-3){10}}\end{picture}}
\put(160,0){\begin{picture}(40,40)
\put(0,0){\circle*{4}}
\put(40,0){\circle*{4}}
\put(20,30){\circle*{4}}
\put(0,0){\vector(1,0){22.5}}
\put(20,0){\line(1,0){20}}
\put(40,0){\vector(-2,3){11.5}}
\put(30,15){\line(-2,3){10}}
\put(20,30){\vector(-2,-3){11.5}}
\put(10,15){\line(-2,-3){10}}\end{picture}}
\put(140,30){\begin{picture}(40,40)
\put(0,0){\circle*{4}}
\put(40,0){\circle*{4}}
\put(20,30){\circle*{4}}
\put(0,0){\vector(1,0){22.5}}
\put(20,0){\line(1,0){20}}
\put(40,0){\vector(-2,3){11.5}}
\put(30,15){\line(-2,3){10}}
\put(20,30){\vector(-2,-3){11.5}}
\put(10,15){\line(-2,-3){10}}\end{picture}}
\put(120,60){\begin{picture}(40,40)
\put(0,0){\circle*{4}}
\put(40,0){\circle*{4}}
\put(20,30){\circle*{4}}
\put(0,0){\vector(1,0){22.5}}
\put(20,0){\line(1,0){20}}
\put(40,0){\vector(-2,3){11.5}}
\put(30,15){\line(-2,3){10}}
\put(20,30){\vector(-2,-3){11.5}}
\put(10,15){\line(-2,-3){10}}\end{picture}}
\put(100,90){\begin{picture}(40,40)
\put(0,0){\circle*{4}}
\put(40,0){\circle*{4}}
\put(20,30){\circle*{4}}
\put(0,0){\vector(1,0){22.5}}
\put(20,0){\line(1,0){20}}
\put(40,0){\vector(-2,3){11.5}}
\put(30,15){\line(-2,3){10}}
\put(20,30){\vector(-2,-3){11.5}}
\put(10,15){\line(-2,-3){10}}\end{picture}}
\put(80,120){\begin{picture}(40,40)
\put(0,0){\circle*{4}}
\put(40,0){\circle*{4}}
\put(20,30){\circle*{4}}
\put(0,0){\vector(1,0){22.5}}
\put(20,0){\line(1,0){20}}
\put(40,0){\vector(-2,3){11.5}}
\put(30,15){\line(-2,3){10}}
\put(20,30){\vector(-2,-3){11.5}}
\put(10,15){\line(-2,-3){10}}\end{picture}}
\put(160,0){\line(-2,3){80}}
\put(160,0){\vector(-2,3){11.5}}
\put(140,30){\vector(-2,3){11.5}}
\put(120,60){\vector(-2,3){11.5}}
\put(100,90){\vector(-2,3){11.5}}

\dashline[10]{4}(80,0)(160,0)
\dashline[10]{4}(60,30)(140,30)
\dashline[10]{4}(40,60)(120,60)
\dashline[10]{4}(60,90)(120,90)
\dashline[10]{4}(80,0)(120,60)
\dashline[10]{4}(60,30)(100,90)
\dashline[10]{4}(40,60)(80,120)
\dashline[10]{4}(120,0)(140,30)
\dashline[10]{4}(120,0)(60,90)

\put(-5,-10){\makebox(0,0){(0,0)}}
\put(40,-10){\makebox(0,0){(1,0)}}
\put(5,30){\makebox(0,0){(0,1)}}
\put(200,-10){\makebox(0,0){$(k,0)$}}
\put(100,160){\makebox(0,0){$(0,k)$}}

\end{picture}
\end{center}
\caption{The ${\cal A}_k$ diagram for $su(3)$ with $(k+1)(k+2)/2$ vertices.}
\label{Akdiagram}
\end{figure}
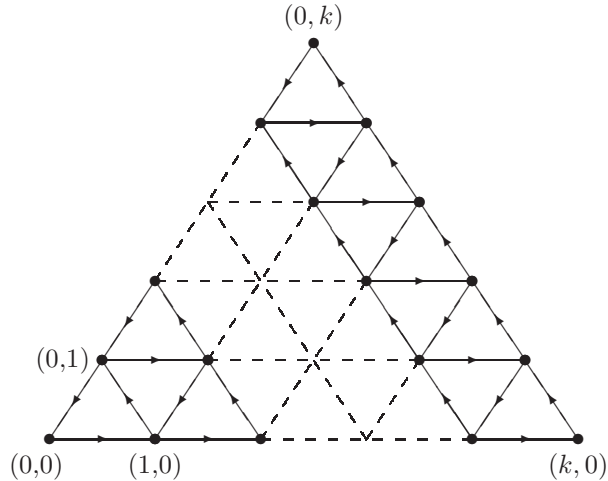

As recalled earlier, this system contains the principal
${\cal A}$ series and three genuine exceptional cases:
${\cal E}_5$,  ${\cal E}_9$ and  ${\cal E}_{21}$.
     The other diagrams of this system  (and in particular the four other
exceptional ones)
     are obtained as twists or as orbifolds of the former list (the
     ``genuine graphs'') , or by  using conjugation and twisting on the
genuine graphs or on their orbifolds.
     A member of the ${\cal A}$ series (a Weyl alcove) is obtained by
truncation of the
     diagram (Weyl chamber) of  tensorisation of irreps of $su(3)$ by one of the
two -- conjugate -- fundamentals $3$ or $\overline 3$; for this
reason the graphs are oriented  (see Fig \ref{Akdiagram}).

The index refers to the level $k$ of the
graph {\sl defined} by  $k  =  \kappa - h = \kappa -3$. Here $h=3$
is the Coxeter number of {\sl the group} $SU(3)$ and $\kappa$
is the generalized Coxeter number of {\sl the graph} (also called
``altitude'').

We label the vertices  $j$ of the ${\cal A}_\kappa$ diagram as
$(\lambda_1, \lambda_2)$, with  $\lambda_1, \lambda_2 \geq 0$
and $\lambda_1 + \lambda_2 \leq k$.
Warning: our labels start from $0$ and not from $1$; many authors
follow a different convention.
Diagrams ${\cal A}_{k}$ have $r$ points with $r=(k+1)(k+2)/2$.

The action of the modular matrix $T$ on vertices $\tau_{j} \equiv
\tau_{(\lambda_{1},
\lambda_{2})}$ of ${\cal A}_{k}$  is diagonal and given by:
$$
\left( T^{(k)}\right)_{\lambda \mu} = e_{\kappa} \left[ -(\lambda_1+1)^2
- (\lambda_1+1).(\lambda_2+1) - (\lambda_2+1)^2 + \kappa \right]
\delta_{\lambda \mu},
$$
where $\lambda \doteq (\lambda_1,\lambda_2)$, $\mu \doteq (\mu_1,\mu_2)$, $e_{\kappa}[x] \doteq
\exp \left( \frac{-2 i \pi x}{3\kappa} \right)$, and $\kappa=k +3$.
We call ``modular exponent'' the quantity $\hat{T} =  -(\lambda_1+1)^2
- (\lambda_1+1).(\lambda_2+1) - (\lambda_2+1)^2 + \kappa $ mod $3 \kappa$.

For $su(3)$, the recurrence formula for adjacency matrices $N_{i}$ associated
with irreps is
$$
\begin{array}{lcll}
N_{\lambda, \mu} &=& 0 & {\text{if} \,   \lambda < 0 \;\, \text{or} \, \mu <0 } \\
N_{\lambda, 0} &=& N_{1,0} N_{\lambda-1 , 0} - N_{\lambda-2 , 1} & {} \\
N_{\lambda, \mu} &=& N_{1,0} N_{\lambda-1 , \mu} -  N_{\lambda-1 ,
\mu - 1} -  N_{\lambda-2 ,\mu + 1} & {\text{if} \, \mu \neq 0} \\
N_{0, \lambda} &=& N_{\lambda, 0}^T & {}
\end{array}
$$
Remember that fused adjacency matrices $F_{i}$, associated with any graph
$G$ of the same level, are  determined by the same recurrence relations
(but the seed is different: $F_{1} = G_{1}$, the adjacency matrix of $G$).

In some cases the vector space generated by the vertices of a
Di Francesco - Zuber graph is an algebra with positive integral
structure constants (self fusion). In all cases it is
a module over the algebra of type ${\cal A}$ with the same Coxeter number.
For an ${\cal A}$ graph, the identity element is  $(0,0)$; vertices
$(1,0)$ and $(0,1)$, corresponding classically to the two
representations of dimension $3$, are the two complex conjugated generators.
As always, a given diagram encodes the multiplication by the
generators in the following
sense: multiplication of an irrep $(\lambda_1, \lambda_2)$ by the
left generator $(1,0)$
is given by the sum of the irreps which are connected to $(\lambda_1,
\lambda_2)$ by an incoming arrow, whereas multiplication by the right
generator $(0,1)$ is given by the sum of the irreps which are
connected to $(\lambda_1, \lambda_2)$ by an outgoing arrow.
To label vertices, some readers may prefer Young frames (diagrams)
rather than a notation using weights.
     The correspondance is as follows:
$(\lambda_1,\lambda_2)$ correspond to Young diagrams $Y(p=\lambda_1 +
\lambda_2, q = \lambda_2)$ with two rows, $p$ boxes on the
first row, and $q$ boxes on the second row.
Graphs whose vector space possesses self fusion have a unit, and one of the two
generators is located at the extremity of the (single) oriented
edge that leaves the origin (reverse the arrows to get the other generator).
Triality \ie \, $\{ 0,1,2 \in \ZZ/3\ZZ\} $ is well defined and
compatible with internal multiplication (if it exists) or
with external multiplication by vertices of the corresponding ${\cal
A}$ graph; it is represented by different choices
of ``colors'' of vertices on the pictures.
There is also a conjugacy transformation  $\sigma \rightarrow
\sigma^c$, at the level of graph matrices, it corresponds to
transposition. For ${\cal A}$ graphs, it is represented by symmetry
with respect to the inner bissectrix of the graph.
The adjacency matrix is not symmetric, but it is normal,
so that it can always be diagonalized.

In the following we illustrate the construction of the Ocneanu graphs
of quantum symmetries, using our method
based on the eigenvalues of the $T$ operator, for the three genuine
exceptional cases. Going through the whole list of
Di Francesco -Zuber graphs would constitute a giant outgrowth of this paper...
We shall give some more details on the ${\cal E}_5$ case than on the
two others.

Notice that the genuine diagrams ${\cal E}_5, {\cal E}_9$ and ${\cal E}_{21}$ are the only ones
among exceptionals to admit self-fusion; this was first observed in \cite{oldZuber}.

\subsection{First example: the ${\cal E}_5$ case}
The ${\cal E}_5$ diagram is illustrated in Figure 8, together with the
corresponding ${\cal A}_5$ diagram, with same norm, equal to $1 + \sqrt
2$, since the altitude is $\kappa = 8$.
Their respective adjacency matrices $G_{1}$ and $N_{1}$ are immediately
determined (the adjacency matrix given in \cite{CFTbook} is not typed correctly).

\begin{figure}[hhh]
\begin{center}
\unitlength 0.30mm

\begin{picture}(440,190)(0,-15)

\put(0,45){\begin{picture}(60,45)
\put(0,0){\color{green} \circle*{5}}
\put(0,0){\color{green} \circle{9}}
\put(60,0){\color{blue} \circle*{5}}
\put(30,45){\color{red} \circle*{5}}
\put(0,0){\vector(1,0){32.5}}
\put(30,0){\line(1,0){30}}
\put(60,0){\vector(-2,3){16.5}}
\put(45,22.5){\line(-2,3){15}}
\put(30,45){\vector(-2,-3){16.5}}
\put(15,22.5){\line(-2,-3){15}}
\end{picture}}

\put(120,45){\begin{picture}(60,45)
\put(0,0){\color{green} \circle*{5}}
\put(60,0){\color{blue} \circle{9}}
\put(60,0){\color{blue} \circle*{5}}
\put(30,45){\color{red} \circle*{5}}
\put(0,0){\vector(1,0){32.5}}
\put(30,0){\line(1,0){30}}
\put(60,0){\vector(-2,3){16.5}}
\put(45,22.5){\line(-2,3){15}}
\put(30,45){\vector(-2,-3){16.5}}
\put(15,22.5){\line(-2,-3){15}}
\end{picture}}

\put(60,135){\begin{picture}(60,45)
\put(0,0){\color{green} \circle*{5}}
\put(60,0){\color{blue} \circle*{5}}
\put(30,45){\color{red} \circle*{5}}
\put(30,45){\color{red} \circle{9}}
\put(0,0){\vector(1,0){32.5}}
\put(30,0){\line(1,0){30}}
\put(60,0){\vector(-2,3){16.5}}
\put(45,22.5){\line(-2,3){15}}
\put(30,45){\vector(-2,-3){16.5}}
\put(15,22.5){\line(-2,-3){15}}
\end{picture}}

\put(0,90){\begin{picture}(60,45)
\put(0,45){\color{blue} \circle*{5}}
\put(0,45){\color{blue} \circle{9}}
\put(60,45){\vector(-1,0){32.5}}
\put(30,45){\line(-1,0){30}}
\put(30,0){\vector(2,3){16.5}}
\put(45,22.5){\line(2,3){15}}
\put(0,45){\vector(2,-3){16.5}}
\put(15,22.5){\line(2,-3){15}}
\end{picture}}

\put(120,90){\begin{picture}(60,45)
\put(60,45){\color{green} \circle*{5}}
\put(60,45){\color{green} \circle{9}}
\put(60,45){\vector(-1,0){32.5}}
\put(30,45){\line(-1,0){30}}
\put(30,0){\vector(2,3){16.5}}
\put(45,22.5){\line(2,3){15}}
\put(0,45){\vector(2,-3){16.5}}
\put(15,22.5){\line(2,-3){15}}
\end{picture}}

\put(60,0){\begin{picture}(60,45)
\put(30,0){\color{red} \circle*{5}}
\put(30,0){\color{red} \circle{9}}
\put(60,45){\vector(-1,0){32.5}}
\put(30,45){\line(-1,0){30}}
\put(30,0){\vector(2,3){16.5}}
\put(45,22.5){\line(2,3){15}}
\put(0,45){\vector(2,-3){16.5}}
\put(15,22.5){\line(2,-3){15}}
\end{picture}}

\put(60,135){\vector(0,-1){47.5}}
\put(60,90){\line(0,-1){45}}
\put(60,45){\vector(2,1){47.2}}
\put(105,67.5){\line(2,1){45}}
\put(150,90){\vector(-2,1){47.2}}
\put(105,112.5){\line(-2,1){45}}

\put(120,45){\vector(0,1){47.5}}
\put(120,90){\line(0,1){45}}
\put(120,135){\vector(-2,-1){47.2}}
\put(75,112.5){\line(-2,-1){45}}
\put(30,90){\vector(2,-1){47.2}}
\put(75,67.5){\line(2,-1){45}}

\put(240,10){\begin{picture}(200,180)

\put(0,0){\begin{picture}(40,40)
\put(0,0){\color{green} \circle*{5}}
\put(40,0){\color{blue} \circle*{5}}
\put(20,30){\color{red} \circle*{5}}
\put(0,0){\vector(1,0){21}}
\put(20,0){\line(1,0){20}}
\put(40,0){\vector(-2,3){11.5}}
\put(30,15){\line(-2,3){10}}
\put(20,30){\vector(-2,-3){11.5}}
\put(10,15){\line(-2,-3){10}}\end{picture}}

\put(40,0){\begin{picture}(40,40)
\put(40,0){\color{red} \circle*{5}}
\put(20,30){\color{green} \circle*{5}}
\put(0,0){\vector(1,0){21}}
\put(20,0){\line(1,0){20}}
\put(40,0){\vector(-2,3){11.5}}
\put(30,15){\line(-2,3){10}}
\put(20,30){\vector(-2,-3){11.5}}
\put(10,15){\line(-2,-3){10}}\end{picture}}

\put(80,0){\begin{picture}(40,40)
\put(40,0){\color{green} \circle*{5}}
\put(20,30){\color{blue} \circle*{5}}
\put(0,0){\vector(1,0){21}}
\put(20,0){\line(1,0){20}}
\put(40,0){\vector(-2,3){11.5}}
\put(30,15){\line(-2,3){10}}
\put(20,30){\vector(-2,-3){11.5}}
\put(10,15){\line(-2,-3){10}}\end{picture}}

\put(120,0){\begin{picture}(40,40)
\put(40,0){\color{blue} \circle*{5}}
\put(20,30){\color{red} \circle*{5}}
\put(0,0){\vector(1,0){21}}
\put(20,0){\line(1,0){20}}
\put(40,0){\vector(-2,3){11.5}}
\put(30,15){\line(-2,3){10}}
\put(20,30){\vector(-2,-3){11.5}}
\put(10,15){\line(-2,-3){10}}\end{picture}}

\put(160,0){\begin{picture}(40,40)
\put(40,0){\color{red} \circle*{5}}
\put(20,30){\color{green} \circle*{5}}
\put(0,0){\vector(1,0){21}}
\put(20,0){\line(1,0){20}}
\put(40,0){\vector(-2,3){11.5}}
\put(30,15){\line(-2,3){10}}
\put(20,30){\vector(-2,-3){11.5}}
\put(10,15){\line(-2,-3){10}}\end{picture}}

\put(20,30){\begin{picture}(40,40)
\put(20,30){\color{blue} \circle*{5}}
\put(0,0){\vector(1,0){21}}
\put(20,0){\line(1,0){20}}
\put(40,0){\vector(-2,3){11.5}}
\put(30,15){\line(-2,3){10}}
\put(20,30){\vector(-2,-3){11.5}}
\put(10,15){\line(-2,-3){10}}\end{picture}}

\put(60,30){\begin{picture}(40,40)
\put(20,30){\color{red} \circle*{5}}
\put(0,0){\vector(1,0){21}}
\put(20,0){\line(1,0){20}}
\put(40,0){\vector(-2,3){11.5}}
\put(30,15){\line(-2,3){10}}
\put(20,30){\vector(-2,-3){11.5}}
\put(10,15){\line(-2,-3){10}}\end{picture}}

\put(100,30){\begin{picture}(40,40)
\put(20,30){\color{green} \circle*{5}}
\put(0,0){\vector(1,0){21}}
\put(20,0){\line(1,0){20}}
\put(40,0){\vector(-2,3){11.5}}
\put(30,15){\line(-2,3){10}}
\put(20,30){\vector(-2,-3){11.5}}
\put(10,15){\line(-2,-3){10}}\end{picture}}

\put(140,30){\begin{picture}(40,40)
\put(20,30){\color{blue} \circle*{5}}
\put(0,0){\vector(1,0){21}}
\put(20,0){\line(1,0){20}}
\put(40,0){\vector(-2,3){11.5}}
\put(30,15){\line(-2,3){10}}
\put(20,30){\vector(-2,-3){11.5}}
\put(10,15){\line(-2,-3){10}}\end{picture}}

\put(40,60){\begin{picture}(40,40)
\put(20,30){\color{green} \circle*{5}}
\put(0,0){\vector(1,0){21}}
\put(20,0){\line(1,0){20}}
\put(40,0){\vector(-2,3){11.5}}
\put(30,15){\line(-2,3){10}}
\put(20,30){\vector(-2,-3){11.5}}
\put(10,15){\line(-2,-3){10}}\end{picture}}

\put(80,60){\begin{picture}(40,40)
\put(20,30){\color{blue} \circle*{5}}
\put(0,0){\vector(1,0){21}}
\put(20,0){\line(1,0){20}}
\put(40,0){\vector(-2,3){11.5}}
\put(30,15){\line(-2,3){10}}
\put(20,30){\vector(-2,-3){11.5}}
\put(10,15){\line(-2,-3){10}}\end{picture}}

\put(120,60){\begin{picture}(40,40)
\put(20,30){\color{red} \circle*{5}}
\put(0,0){\vector(1,0){21}}
\put(20,0){\line(1,0){20}}
\put(40,0){\vector(-2,3){11.5}}
\put(30,15){\line(-2,3){10}}
\put(20,30){\vector(-2,-3){11.5}}
\put(10,15){\line(-2,-3){10}}\end{picture}}

\put(60,90){\begin{picture}(40,40)
\put(20,30){\color{red} \circle*{5}}
\put(0,0){\vector(1,0){21}}
\put(20,0){\line(1,0){20}}
\put(40,0){\vector(-2,3){11.5}}
\put(30,15){\line(-2,3){10}}
\put(20,30){\vector(-2,-3){11.5}}
\put(10,15){\line(-2,-3){10}}\end{picture}}

\put(100,90){\begin{picture}(40,40)
\put(20,30){\color{green} \circle*{5}}
\put(0,0){\vector(1,0){21}}
\put(20,0){\line(1,0){20}}
\put(40,0){\vector(-2,3){11.5}}
\put(30,15){\line(-2,3){10}}
\put(20,30){\vector(-2,-3){11.5}}
\put(10,15){\line(-2,-3){10}}\end{picture}}

\put(80,120){\begin{picture}(40,40)
\put(20,30){\color{blue} \circle*{5}}
\put(0,0){\vector(1,0){21}}
\put(20,0){\line(1,0){20}}
\put(40,0){\vector(-2,3){11.5}}
\put(30,15){\line(-2,3){10}}
\put(20,30){\vector(-2,-3){11.5}}
\put(10,15){\line(-2,-3){10}}\end{picture}}

\put(-5,-10){\makebox(0,0){(0,0)}}
\put(40,-10){\makebox(0,0){(1,0)}}
\put(3,30){\makebox(0,0){(0,1)}}
\put(200,-10){\makebox(0,0){$(5,0)$}}
\put(100,160){\makebox(0,0){$(0,5)$}}

\end{picture}}

\put(-12,45){\makebox(0,0){$1_0$}}
\put(192,45){\makebox(0,0){$1_4$}}
\put(-12,135){\makebox(0,0){$1_1$}}
\put(192,135){\makebox(0,0){$1_3$}}
\put(90,-12){\makebox(0,0){$1_5$}}
\put(90,192){\makebox(0,0){$1_2$}}

\put(55,35){\makebox(0,0){$2_1$}}

\put(125,35){\makebox(0,0){$2_0$}}
\put(55,145){\makebox(0,0){$2_3$}}
\put(125,145){\makebox(0,0){$2_4$}}
\put(20,90){\makebox(0,0){$2_2$}}
\put(160,90){\makebox(0,0){$2_5$}}

\end{picture}

\caption{The ${\cal E}_5$ and ${\cal A}_5$ generalized Dynkin diagrams}

\end{center}
\end{figure}
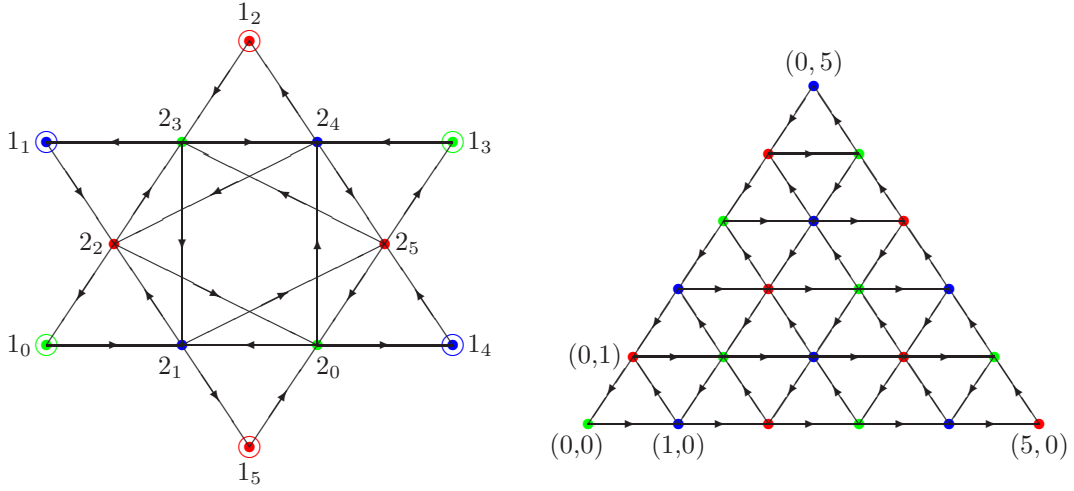

The ${\cal E}_5$ diagram admits self-fusion; $1_0$ is the identity, $2_1$ and $2_2$
are the left and right generators.
The multiplication table of the graph algebra of ${\cal E}_5$ reads:
$$
\begin{array}{ccl}
1_j . 1_k & = & 1_{j+k}   \\
1_j . 2_k & = & 2_{j+k}   \\
2_j . 2_k & = & 2_{j+k} + 2_{j+k-3} + 1_{j+k-3}
\end{array}
$$
This multiplication table allows one to compute easily the $12$ square
matrices $G_{a}$ of the graph. The subset $1_i$ clearly forms a subalgebra
of the graph algebra.

\subsubsection{Restriction mechanism}
We define an action of ${\cal A}_5$ on ${\cal E}_5$ in the same way as
for the previous cases (see the discussion
for $E_6$), getting
the following restrictions: $(0,0) \hookrightarrow 1_0$, $(1,0)
\hookrightarrow 2_1$ and $(0,1) \hookrightarrow 2_2$.
For the others points, we compute the powers
$(1,0)^{\alpha}(0,1)^{\beta}$ of the two fundamentals as well as
the powers $(2_1)^{\alpha}(2_2)^{\beta}$ and compare them:
$$
\begin{array}{lll}
(1,0)^2 = (2,0) + (0,1), & (2_1)^2 = 2_2 + 2_5 + 1_5, & \textrm{so}
\quad (2,0) \hookrightarrow 1_5 + 2_5; \\
(0,1)^2 = (1,0) + (0,2), & (2_2)^2 = 2_1 + 2_4 + 1_1, & \textrm{so}
\quad (0,2) \hookrightarrow 1_1 + 2_4; \\
(1,0).(0,1) = (0,0) + (1,1), & 2_1 . 2_2 = 1_0 + 2_0 + 2_3, & \textrm{so}
\quad (1,1) \hookrightarrow 2_0 + 2_3; \\
(1,0)^3 = 2(1,1) + (3,0) + (0,0), & (2_1)^3 = 1_0 + 3 2_0 + 2  2_3 +
1_3, & \textrm{so} \quad (3,0) \hookrightarrow 2_0 + 1_3;
\end{array}
$$
and so on...

 From these restriction rules, we obtain immediately the lines of essential
matrix $E_{0}$(intertwiner): it is a rectangular matrix with $12$ columns, indexed
by vertices of ${\cal E}_{5}$ and $21$ rows indexed by vertices of ${\cal
A}_{5}$ (\ie by pairs of integers $(\lambda_{1}, \lambda_{2})$ with
$\lambda_{1} +
\lambda_{2} \leq 5$ or by Young frames $Y(p,q)$ with $5 \geq p \geq q$).

We could have, as well, calculated directly the $21$ fused matrices
$F_{i}$ from $G_{1}$ alone by using the $su(3)$ recurrence relations; these
matrices, in turn, determine the $12$ essential (rectangular) matrices $E_{a}$.

\begin{figure}[h]
\begin{center}
\includegraphics*{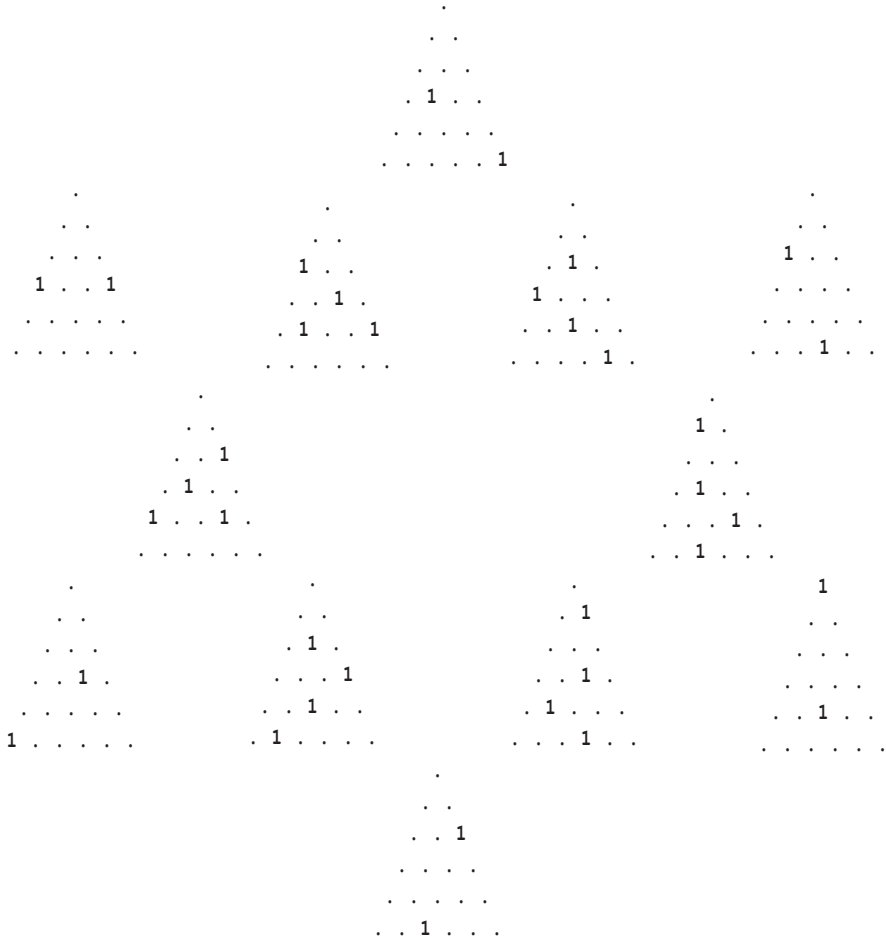}
\caption{Matrix $E_0$ for ${\cal E}_5$}
\label{E5E0mat}
\end{center}
\end{figure}

\subsubsection{Induction mechanism}
 From the branching rules ${\cal A}_5 \rightarrow {\cal E}_5$, 
we get the following induction rules:
$$
\begin{array}{ccc}
1_0 \hookleftarrow (0,0) , (2,2) &\qquad& 2_1 \hookleftarrow (1,0) ,
(2,1) , (1,3) , (3,2) \\
1_1 \hookleftarrow (0,2) , (3,2) &\qquad& 2_2 \hookleftarrow (0,1) ,
(1,2) , (3,1) , (2,3) \\
1_2 \hookleftarrow (1,2) , (5,0) &\qquad& 2_3 \hookleftarrow (1,1) ,
(0,3) , (2,2) , (4,1) \\
1_3 \hookleftarrow (3,0) , (0,3) &\qquad& 2_4 \hookleftarrow (0,2) ,
(2,1) , (4,0) , (1,3) \\
1_4 \hookleftarrow (2,1) , (0,5) &\qquad& 2_5 \hookleftarrow (2,0) ,
(1,2) , (3,1) , (0,4) \\
1_5 \hookleftarrow (2,0) , (2,3) &\qquad& 2_0 \hookleftarrow (1,1) ,
(3,0) , (2,2) , (1,4)
\end{array}
$$
The same information can be gathered from the columns of  matrix $E_0$ (see Fig \ref{E5E0mat}:
each triangle corresponds to a single column).
The first rule can be interpreted as a manifestation of the existence
of a non trivial quantum invariant of ``degree'' $(2,2)$.

\subsubsection{Quantum symmetries}
For each $1_i$, we can verify that the values of $T$ on the two
corresponding $(\lambda_1, \lambda_2)$ coming
from the induction are the same. This allows us to assign a fixed
value of T to the $1_i$'s. We can
also verify that we can not do the same for the other vertices
$2_i$'s. We get in this way a characterization
of the subalgebra $J$, spanned by the elements $1_i$'s.
\begin{table}
\small
$$
\begin{array}{|c||c|c|c|c|c|c|c|c|c|c|c|c|}
\hline
(\lambda_1,\lambda_2) & (0,0) & (1,0) &(2,0) &(3,0) &(4,0) &(5,0)
&(1,1) &(2,1) &(3,1) &(4,1) &(2,2) &(3,2) \\
                        &  {}   & (0,1) &(0,2) &(0,3) &(0,4) &(0,5) &
{}   &(1,2) &(1,3) &(1,4) & {}   &(2,3) \\
\hline
\hline
\hat{T}  & 5 & 1 & 19 & 11 & 1 & 13 & 20 & 13 & 4 & 17 & 5 & 19 \\
\hline
\end{array}
$$
\normalsize
\caption{Values of $\hat{T}$ on the vertices of the $\mathcal{A}_5$ graph}
\end{table}

\begin{figure}
\begin{center}
\includegraphics*{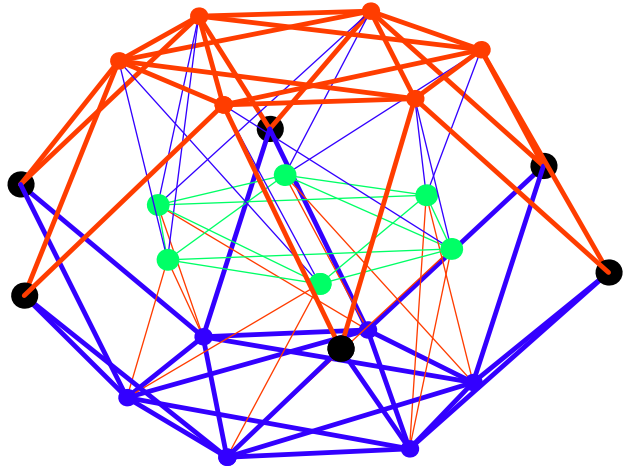}
\caption{Ocneanu graph of ${\cal E}_{5}$}
\label{OcGraphE5}
\end{center}
\end{figure}

We therefore expect the algebra of quantum symmetries of ${\cal E}_{5}$
to be $Oc({\cal E}_{5} ) = {\cal E}_{5}  \otimes_{J} {\cal
E}_{5}$. Its dimension is $12.12/6 = 24$.
The left and right subalgebras are
respectively spanned by $L = \{a \otimesdot 1_{0}\}$ and  $R =
\{1_{0} \otimesdot a\}$, with $a$ equal to $2_{j}$ or $1_{j}$. Both
left and right chiral subgraphs
have $12$ points. The ambichiral subalgebra (of dimension $6$) is spanned
by $A = \{1_{j} \otimesdot 1_{0} = 1_{0} \otimesdot 1_{j} \}$ and the
supplementary subspace (also $6$ points) is spanned by  $C = \{2_{j}
\otimesdot 2_{k} =
2_{0} \otimesdot 2_{j+k} \}$. The Ocneanu graph can be displayed on
the (three dimensional) picture (Fig \ref{OcGraphE5}) as two superposed stars kissing
each other along the six ambichiral points, with the vertices spanning the
supplement displayed ``inside'' the others. As usual, bold lines ---
of two different colors ---  refer
to the chiral parts and thin lines to the corresponding quotients. Left chiral graph is blue 
(bold lines); right chiral graph is red (bold lines). Ambichiral points are black and points belonging 
to the suppplementary subspace are green. The action of the left generator $2_1 \otimesdot 1_0$
(right generator $1_0 \otimesdot 2_1$) on any point is a linear combination of blue (red) lines.
Green lines (bold) are understood as both red and blue thin lines. This graph is oriented but we
have not displayed the orientation of the edges in order not to clutter the picture; the
interested reader should do it for himself.

\subsubsection{Dimensions of blocks}

The two multiplicative structures $\circ$ and $\star$ of the bialgebra
${\cal B}{\cal E}_{5}$ can be diagonalized.\\
Blocks corresponding to
the first structure are labelled by the $21$ points of the ${\cal A}_{5}$
diagram. Dimension $d_{i}$ of the block $i$ is obtained by summing
the matrix elements
of $F_{i}$. We order the blocks $i = (\lambda, \mu)$
according to the level, \ie $(\lambda, \mu) < (\lambda', \mu')$ if
$\lambda < \lambda'$ or $\lambda = \lambda'$ and $\mu < \mu'$, and find:
$$
{\{12\}, \{24, 24\}, \{36, 48, 36\}, \{36, 60, 60, 36\}, \{24, 48, 60,
48, 24\}, \{12,
       24, 36, 36, 24, 12\}}
       $$
Dimension of the bialgebra is obtained by summing the square of these
$21$ integers $d_{i}$: $dim ({\cal B}{\cal E}_{5}) = 29376$.
Dimension of the vector space of essential paths (graded by the
Young frames of  ${\cal A}_{5}$) is $\sum_{i} d_{i} = 720$

Blocks corresponding to
the second structure are labelled by the $24$ points of the Ocneanu
graph $Oc( {\cal E}_{5})$. Dimension $d_{x}$ of the block $x$
is obtained by summing the matrix elements
of matrices $S_{x} = G_{a} G_{b}$ when $x = a \otimesdot b$ runs over
the points of $Oc( {\cal E}_{5})$. One finds the following:
the $6$ ambichiral blocks have dimension $12$, the six left chiral
and the six right chiral blocks which are not ambichiral have dimension
$24$, the six complementary blocks have dimension $60$.
Dimension of the bialgebra is also obtained by summing the square of
these $24$ integers $d_{x}$ and one finds the same total as before.
Notice that writing $29376$ in two different ways as a sum of $21$ or
$24$ squares
constitutes, of course, a rather non trivial check.
Notice that we find also  $\sum_x d_{x} = 720$.

We summarize the discussion as follows: 
$$
dim({\cal B}{\cal E}_5) = 
\sum_{i \in {\cal A}_{5}} d_{i}^{2} = \sum_{x \in Oc({\cal E}_{5})}
d_{x}^{2} = (2)^{6} (3)^3 (17)^1 \quad \text{and} \quad
\sum_{i\in {\cal A}_{5}} d_{i} =
\sum_{x \in Oc( {\cal E}_{5})} d_{x} =  (2)^{4} (3)^{2} (5)^1
$$

\subsubsection{Toric matrices and twisted partition functions}
    From the essential matrices $E_{a}$, we easily calculate the toric
matrices (square matrices of dimension $21$) and the corresponding
partition functions by the method described
earlier.  There is one such function for each point of the Ocneanu
graph $Oc({\cal E}_5)$.  The one obtained from the identity $1_{0} 
\otimesdot 1_{0}$
of the graph is the modular-invariant and agrees with the expression of \cite{Gannon}
(there is a global shift of $(1,1)$ due to our conventions):
$$
\begin{array}{lcc}
{\cal Z}_{{\cal E}_5}\doteq {\cal Z}_{1_0 \otimesdot 1_0} &=& | \chi_{(0,0)} + \chi_{(2,2)} |^2 + |
\chi_{(0,2)} + \chi_{(3,2)} |^2 +
| \chi_{(2,0)} + \chi_{(2,3)} |^2  \\
{} &+& | \chi_{(2,1)} + \chi_{(0,5)} |^2 + | \chi_{(3,0)} + \chi_{(0,3)} |^2
+ | \chi_{(1,2)} + \chi_{(5,0)} |^2
\end{array}
$$

The others are interpreted as twisted partition functions (one defect
line, in the interpretation of \cite{PetZub:Oc}). 
We give only the twisted partition functions associated with ambichiral points 
$1_{0} \otimesdot 1_{i}$, for $i \in \{1,2,3,4,5\}$:
\scriptsize
$$
\begin{array}{ccl}
{\cal Z}_{1_0\otimesdot1_1} &=& (\chi_{(0, 3)} + \chi_{(3,0)}).(\ov{\chi}_{(0, 5)} + \ov{\chi}_{2,1)}) 
        +  (\chi_{(2, 0)} + \chi_{(2,3)}).(\ov{\chi}_{(0, 0)} + \ov{\chi}_{2,2)}) 
        +  (\chi_{(0, 5)} + \chi_{(2,1)}).(\ov{\chi}_{(2, 0)} + \ov{\chi}_{2,3)}) \\
       &+& (\chi_{(1, 2)} + \chi_{(5,0)}).(\ov{\chi}_{(0, 3)} + \ov{\chi}_{3,0)}) 
        +  (\chi_{(0, 0)} + \chi_{(2,2)}).(\ov{\chi}_{(0, 2)} + \ov{\chi}_{3,2)}) 
        +  (\chi_{(0, 2)} + \chi_{(3,2)}).(\ov{\chi}_{(1, 2)} + \ov{\chi}_{5, 0)})\\
{} & {} & {} \\
{\cal Z}_{1_0\otimesdot1_2} &=& (\chi_{(1, 2)} + \chi_{(5,0)}).(\ov{\chi}_{(0, 5)} + \ov{\chi}_{2,1)}) 
        +  (\chi_{(0, 5)} + \chi_{(2,1)}).(\ov{\chi}_{(0, 0)} + \ov{\chi}_{2,2)}) 
        +  (\chi_{(0, 3)} + \chi_{(3,0)}).(\ov{\chi}_{(2, 0)} + \ov{\chi}_{2,3)}) \\
       &+& (\chi_{(0, 2)} + \chi_{(3,2)}).(\ov{\chi}_{(0, 3)} + \ov{\chi}_{3,0)}) 
        +  (\chi_{(2, 0)} + \chi_{(2,3)}).(\ov{\chi}_{(0, 2)} + \ov{\chi}_{3,2)}) 
        +  (\chi_{(0, 0)} + \chi_{(2,2)}).(\ov{\chi}_{(1, 2)} + \ov{\chi}_{5, 0)}) \\
{} & {} & {} \\
{\cal Z}_{1_0\otimesdot1_3} &=& (\chi_{(0, 2)} + \chi_{(3,2)}).(\ov{\chi}_{(0, 5)} + \ov{\chi}_{2,1)}) 
        +  (\chi_{(0, 3)} + \chi_{(3,0)}).(\ov{\chi}_{(0, 0)} + \ov{\chi}_{2,2)}) 
        +  (\chi_{(1, 2)} + \chi_{(5,0)}).(\ov{\chi}_{(2, 0)} + \ov{\chi}_{2,3)})  \\
       &+& (\chi_{(0, 0)} + \chi_{(2,2)}).(\ov{\chi}_{(0, 3)} + \ov{\chi}_{3,0)}) 
        +  (\chi_{(0, 5)} + \chi_{(2,1)}).(\ov{\chi}_{(0, 2)} + \ov{\chi}_{3,2)}) 
        +  (\chi_{(2, 0)} + \chi_{(2,3)}).(\ov{\chi}_{(1, 2)} + \ov{\chi}_{5, 0)}) \\
{} & {} & {} \\
{\cal Z}_{1_0\otimesdot1_4} &=& (\chi_{(0, 0)} + \chi_{(2,2)}).(\ov{\chi}_{(0, 5)} + \ov{\chi}_{2,1)}) 
        +  (\chi_{(1, 2)} + \chi_{(5,0)}).(\ov{\chi}_{(0, 0)} + \ov{\chi}_{2,2)}) 
        +  (\chi_{(0, 2)} + \chi_{(3,2)}).(\ov{\chi}_{(2, 0)} + \ov{\chi}_{2,3)}) \\
       &+& (\chi_{(2, 0)} + \chi_{(2,3)}).(\ov{\chi}_{(0, 3)} + \ov{\chi}_{3,0)}) 
        +  (\chi_{(0, 3)} + \chi_{(3,0)}).(\ov{\chi}_{(0, 2)} + \ov{\chi}_{3,2)}) 
        +  (\chi_{(0, 5)} + \chi_{(2,1)}).(\ov{\chi}_{(1, 2)} + \ov{\chi}_{5, 0)}) \\
{} & {} & {} \\
{\cal Z}_{1_0\otimesdot1_5} &=& (\chi_{(2, 0)} + \chi_{(2,3)}).(\ov{\chi}_{(0, 5)} + \ov{\chi}_{2,1)}) 
        +  (\chi_{(0, 2)} + \chi_{(3,2)}).(\ov{\chi}_{(0, 0)} + \ov{\chi}_{2,2)}) 
        +  (\chi_{(0, 0)} + \chi_{(2,2)}).(\ov{\chi}_{(2, 0)} + \ov{\chi}_{2,3)}) \\
       &+& (\chi_{(0, 5)} + \chi_{(2,1)}).(\ov{\chi}_{(0, 3)} + \ov{\chi}_{3,0)}) 
        +  (\chi_{(1, 2)} + \chi_{(5,0)}).(\ov{\chi}_{(0, 2)} + \ov{\chi}_{3,2)}) 
        +  (\chi_{(0, 3)} + \chi_{(3,0)}).(\ov{\chi}_{(1, 2)} + \ov{\chi}_{5, 0)})
\end{array}
$$
\normalsize

\subsection{Second example: the ${\cal E}_9$ case}

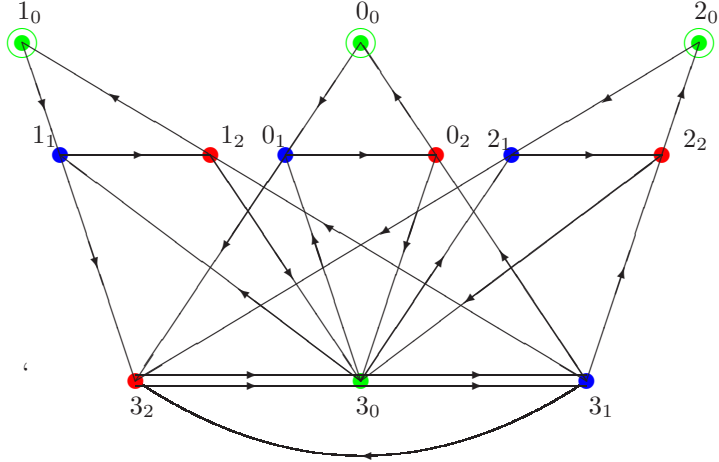
\begin{figure}[hhh]
\unitlength=0.5mm
\begin{center}

\begin{picture}(180,130)(0,-20)
\put(70,60){\begin{picture}(40,30)
\put(0,0){\color{blue} \circle*{4}}
\put(40,0){\color{red} \circle*{4}}
\put(20,30){\color{green} \circle*{4}}
\put(0,0){\vector(1,0){22.5}}
\put(20,0){\line(1,0){20}}
\put(40,0){\vector(-2,3){11.5}}
\put(30,15){\line(-2,3){10}}
\put(20,30){\vector(-2,-3){11.5}}
\put(10,15){\line(-2,-3){10}}
\end{picture}}

\qbezier[920](30,0)(90,-40)(150,0)
\put(90,-20){\vector(-1,0){0}}

\put(30,0){\color{red} \circle*{4}}
\put(90,0){\color{green} \circle*{4}}
\put(150,0){\color{blue} \circle*{4}}
\put(10,60){\color{blue} \circle*{4}}
\put(50,60){\color{red} \circle*{4}}
\put(130,60){\color{blue} \circle*{4}}
\put(170,60){\color{red} \circle*{4}}
\put(0,90){\color{green} \circle*{4}}
\put(180,90){\color{green} \circle*{4}}

\put(0,90){\color{green} \circle{8}}
\put(90,90){\color{green} \circle{8}}
\put(180,90){\color{green} \circle{8}}

\put(0,90){\vector(1,-3){6.0}}
\put(10,60){\line(-1,3){4.5}}
\put(10,60){\vector(1,0){22.5}}
\put(50,60){\line(-1,0){17.5}}

\put(0,90){\line(5,-3){150}}
\put(23,76.4){\vector(-3,2){0}}

\put(30,0){\line(-1,3){10.8}}
\put(10,60){\vector(1,-3){10}}

\put(110,60){\vector(-1,-3){8.6}}
\put(90,0){\line(1,3){15}}
\put(90,0){\vector(-1,3){13}}
\put(70,60){\line(1,-3){10}}

\put(170,60){\line(-1,-3){10.8}}
\put(150,0){\vector(1,3){10}}

\put(180,90){\line(-5,-3){150}}
\put(154,74){\vector(-3,-2){0}}

\put(130,60){\vector(1,0){22.5}}
\put(170,60){\line(-1,0){17.5}}
\put(170,60){\vector(1,3){6.0}}
\put(180,90){\line(-1,-3){4.5}}

\put(70,60){\line(-2,-3){40}}
\put(70,60){\vector(-2,-3){17}}
\put(110,60){\line(2,-3){40}}
\put(150,0){\vector(-2,3){23}}

\put(30,1.5){\line(1,0){120}}
\put(30,-1.5){\line(1,0){120}}
\put(30,1.5){\vector(1,0){32.5}}
\put(30,-1.5){\vector(1,0){32.5}}
\put(90,1.5){\vector(1,0){32.5}}
\put(90,-1.5){\vector(1,0){32.5}}

\put(90,0){\line(-4,3){80}}
\put(90,0){\vector(-4,3){32.5}}
\put(90,0){\line(4,3){80}}
\put(170,60){\vector(-4,-3){52.5}}

\put(50,60){\line(2,-3){40}}
\put(50,60){\vector(2,-3){22}}
\put(90,0){\line(2,3){40}}
\put(90,0){\vector(2,3){22}}

\put(80,41.7){\vector(-3,2){0}}
\put(95,38.7){\vector(-3,-2){0}}
`
\put(0,98){\makebox(0,0){$1_0$}}
\put(90,98){\makebox(0,0){$0_0$}}
\put(180,98){\makebox(0,0){$2_0$}}

\put(4,65){\makebox(0,0){$1_1$}}
\put(65,65){\makebox(0,0){$0_1$}}
\put(125,64){\makebox(0,0){$2_1$}}

\put(54,65){\makebox(0,0){$1_2$}}
\put(114,65){\makebox(0,0){$0_2$}}
\put(177,64){\makebox(0,0){$2_2$}}

\put(30,-7){\makebox(0,0){$3_2$}}
\put(90,-7){\makebox(0,0){$3_0$}}
\put(152,-7){\makebox(0,0){$3_1$}}

\end{picture}
\end{center}
\caption{Diagramme de Dynkin g\'en\'eralis\'e ${\cal E}_9$}
\label{E9diagram}
\end{figure}

This diagram is illustrated on Fig \ref{E9diagram} (notice that it would be 
better drawn three-dimensionally as a small starwars spaceship with 
two wings and a cockpit, because of the existing symmetries between 
the two wings, reminiscent of what happens for the $D_{2n}$ 
Dynkin diagrams).

\begin{figure}
\begin{center}
\includegraphics*{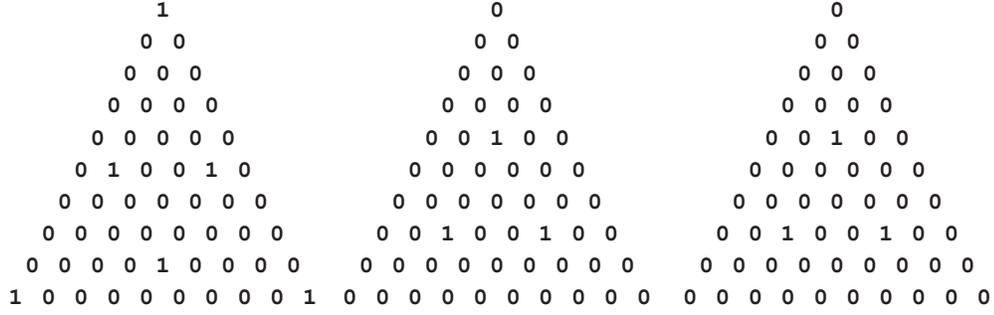}
\caption{Induction corresponding to the three upper vertices of ${\cal E}_9$}
\label{E0E9Col}
\end{center}
\end{figure}

The corresponding diagram of the $A$ series is ${\cal A}_9$. Altitude 
of both is $\kappa = 9+3=12$. Their respective adjacency matrices are 
immediately read from the graphs. Their number of vertices are $12$ 
and $10\times 11/2 = 55$.
Restriction and induction is studied as usual, and imposing constancy 
of the modular operator $T$ singles out the three circled vertices of 
Fig \ref{E9diagram} as elements of the vector subspace $J$ that is used to 
characterize the ambichiral points of the Ocneanu graph. The fused 
adjacency matrices $F_i$ are obtained from the $su(3)$ recurrence 
formula; this determines the essential matrices $E_a$. We give on 
Fig \ref{E0E9Col} the columns of the $E_0$ matrix indexed by the three 
special points (these are the ``ambichiral columns'' of the 
intertwiner $E_0$); a consistent value of $T$ can be defined for 
these three points (and these three points only), one finds $\hat{T} = 9$ 
for the vertex $0_0$ and $\hat{T}=21$ for $1_0$ and $2_0$.

\begin{figure}[hhh]
\begin{center}
\includegraphics*{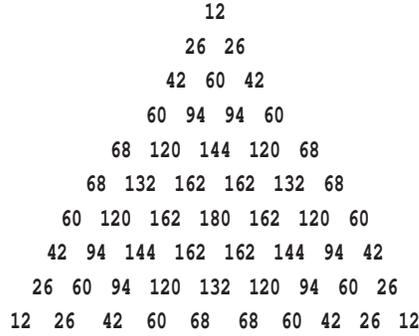}
\caption{Dimension of space of blocks (law $\circ$ ) for $\mathcal{E}_9$}
\label{DimpathsE9}
\end{center}
\end{figure}

Blocks of the bialgebra ${\cal B}{\cal E}_9$, for its first 
associative law ($\circ$), are labelled by the $55$ vertices of ${\cal 
A}_9$ and their dimensions are given on Fig \ref{DimpathsE9}. The total 
dimension is the sum of corresponding squares: $dim({\cal B}
{\cal E}_9) = \sum_i d_i^2 = 518976 = (2)^6(3)^2(17)^1(53)^1 $.

Something special happens however for this graph (again reminiscent 
of a similar situation in the $D_{2n}$ case of Dynkin diagrams): 
first of all, the diagram itself is not sufficient to determine a 
unique associative algebra structure, and one has to impose 
positivity and integrality of the structure constants in order to 
determine a self-fusion structure (it is unique up to permutation of 
the two wings). Since the determination of the corresponding 
graph matrices is not totally straightforward, we give below the 
two matrices corresponding to the endpoints $1_0$ and $2_0$. We 
choose the following order for the vertices : $0_0,1_0,2_0,3_0; 
0_1,1_1,2_1,3_1; 0_2,1_2,2_2,3_2$. We also give the adjacency matrix 
$G_{0_1}$ whose determination {\sl is} straightforward.
\scriptsize
$$
G_{{0}_{1}}  = 
\left( \begin{array}{cccccccccccc}
. & . & . & . & . & . & . & . & 1 & . & . & .  \\
. & . & . & . & . & . & . & . & . & 1 & . & .  \\
. & . & . & . & . & . & . & . & . & . & 1 & .  \\
. & . & . & . & . & . & . & . & 1 & 1 & 1 & 2 \\
1 & . & . & 1 & . & . & . & . & . & . & . & .  \\
. & 1 & . & 1 & . & . & . & . & . & . & . & .  \\
. & . & 1 & 1 & . & . & . & . & . & . & . & .  \\
. & . & . & 2 & . & . & . & . & . & . & . & .  \\
. & . & . & . & 1 & . & . & 1 & . & . & . & .  \\
. & . & . & . & . & 1 & . & 1 & . & . & . & .  \\
. & . & . & . & . & . & 1 & 1 & . & . & . & .  \\
. & . & . & . & 1 & 1 & 1 & 1 & . & . & . & .  \\
\end{array}
\right)
$$

$$
G_{{1}_{0}}  = 
\left( \begin{array}{cccccccccccc}
. & 1 & . & . & . & . & . & . & . & . & . & .  \\
. & . & 1 & . & . & . & . & . & . & . & . & .  \\
1 & . & . & . & . & . & . & . & . & . & . & .  \\
. & . & . & 1 & . & . & . & . & . & . & . & .  \\
. & . & . & . & . & 1 & . & . & . & . & . & .  \\
. & . & . & . & . & . & 1 & . & . & . & . & .  \\
. & . & . & . & 1 & . & . & . & . & . & . & .  \\
. & . & . & . & . & . & . & 1 & . & . & . & .  \\
. & . & . & . & . & . & . & . & . & 1 & . & .  \\
. & . & . & . & . & . & . & . & . & . & 1 & .  \\
. & . & . & . & . & . & . & . & 1 & . & . & .  \\
. & . & . & . & . & . & . & . & . & . & . & 1 
\end{array}
\right)
\quad 
G_{{2}_{0}}  = 
\left( \begin{array}{cccccccccccc}
. & . & 1 & . & . & . & . & . & . & . & . & .  \\
1 & . & . & . & . & . & . & . & . & . & . & .  \\
. & 1 & . & . & . & . & . & . & . & . & . & .  \\
. & . & . & 1 & . & . & . & . & . & . & . & .  \\
. & . & . & . & . & . & 1 & . & . & . & . & .  \\
. & . & . & . & 1 & . & . & . & . & . & . & .  \\
. & . & . & . & . & 1 & . & . & . & . & . & .  \\
. & . & . & . & . & . & . & 1 & . & . & . & .  \\
. & . & . & . & . & . & . & . & . & . & 1 & .  \\
. & . & . & . & . & . & . & . & 1 & . & . & .  \\
. & . & . & . & . & . & . & . & . & 1 & . & .  \\
. & . & . & . & . & . & . & . & . & . & . & 1 
\end{array}
\right)
$$
\normalsize

Next, and as expected, the operator $T$ does not distinguish 
between these two points, and we therefore expect, as in the 
$D_{2n}$ case of the $su(2)$ system, that the algebra $Oc({\cal 
E}_9)$ of quantum symmetries will possess a non-commutative $2\times 
2$ matrix component, encoding, in a ``non-commutative geometrical 
spirit'', this indistinguishability. The presence of such a non 
commutative piece is also reflected in the presence of a coefficient 
$2$ in the (known) modular invariant partition function. We note, 
however, that ambichiral points are bound to be, in any case, $0_0 
\otimesdot 0_0$, $1_0 \otimesdot 0_0 = 0_0 \otimesdot 1_0$ and $2_0 
\otimesdot 0_0 = 0_0 \otimesdot 2_0$. The corresponding toric 
matrices $W$ and partition functions ${\cal Z}$ are computed as usual. We 
define the linear combination $U$ and $V$ of characters:
$$
\begin{array}{rcl}
U &=& \chi_{(2,2)} + \chi_{(2,5)} + \chi_{(5,2)} \\
V &=& \chi_{(0,0)} + \chi_{(0,9)} + \chi_{(9,0)} + \chi_{(1,4)} + \chi_{(4,1)} + \chi_{(4,4)} 
\end{array}
$$
and find:
$$
\begin{array}{rcl}
{\cal Z}_{0_0 \otimesdot 0_0} &=& 2 \, U.\ov{U} + V.\ov{V}   \\
{\cal Z}_{1_0 \otimesdot 0_0} = {\cal Z}_{2_0 \otimesdot 0_0} &=& U.\ov{V} + V.\ov{U} 
\end{array}
$$
The first one is modular invariant and agrees with the expression of 
Gannon \cite{Gannon}. The other one should be interpreted as a twisted 
partition function in a BCFT with defect lines.

Unfortunately, in this case, as it was for $D_{2n}$, the data 
provided by the eigenvalues of the modular operator $T$ does not seem 
to be sufficient to determine the full (non commutative in this case) 
structure of $Oc({\cal E}_9)$  or the Ocneanu graph itself, and we 
decide to stop at this point.

\subsection{Third example: the ${\cal E}_{21}$ case}

The ${\cal E}_{21}$ diagram is illustrated in Fig 9. The
corresponding ${\cal A}$ diagram with same norm is ${\cal A}_{21}$.
The altitude of both is $\kappa = 21 + 3 = 24$.
Their respective adjacency matrices $G_{1}$ and $N_{1}$ are immediately
obtained from the diagrams. The number of vertices of the two diagrams
are respectively equal to $24$ and $22\times 23/2 = 253$.

\begin{figure}[hhh]
\begin{center}
\begin{picture}(320,170)

\put(0,60){\begin{picture}(40,40)
\put(0,20){\color{green} \circle*{4}}
\put(40,0){\color{blue} \circle*{4}}
\put(40,40){\color{red} \circle*{4}}
\put(0,20){\line(2,-1){40}}
\put(0,20){\line(2,1){40}}
\put(40,0){\line(0,1){40}}
\put(40,40){\vector(-2,-1){22.5}}
\put(0,20){\vector(2,-1){22.5}}
\put(40,0){\vector(0,1){22.5}}
\end{picture}}

\put(40,40){\begin{picture}(40,40)
\put(40,0){\color{red} \circle*{4}}
\put(40,40){\color{green} \circle*{4}}
\put(0,20){\line(2,-1){40}}
\put(0,20){\line(2,1){40}}
\put(40,0){\line(0,1){40}}
\put(40,40){\vector(-2,-1){22.5}}
\put(0,20){\vector(2,-1){22.5}}
\put(40,0){\vector(0,1){22.5}}
\end{picture}}

\put(40,80){\begin{picture}(40,40)
\put(40,40){\color{blue} \circle*{4}}
\put(0,20){\line(2,-1){40}}
\put(0,20){\line(2,1){40}}
\put(40,0){\line(0,1){40}}
\put(40,40){\vector(-2,-1){22.5}}
\put(0,20){\vector(2,-1){22.5}}
\put(40,0){\vector(0,1){22.5}}
\end{picture}}

\put(80,60){\begin{picture}(40,40)
\put(40,0){\color{blue} \circle*{4}}
\put(40,40){\color{red} \circle*{4}}
\put(0,20){\line(2,-1){40}}
\put(0,20){\line(2,1){40}}
\put(40,0){\line(0,1){40}}
\put(40,40){\vector(-2,-1){22.5}}
\put(0,20){\vector(2,-1){22.5}}
\put(40,0){\vector(0,1){22.5}}
\end{picture}}

\put(80,40){\line(4,-1){80}}
\put(80,40){\vector(4,-1){22.5}}

\put(80,120){\line(4,1){80}}
\put(160,140){\vector(-4,-1){62.5}}

\put(280,60){\begin{picture}(40,40)
\put(0,0){\color{red} \circle*{4}}
\put(40,20){\color{green} \circle*{4}}
\put(0,40){\color{blue} \circle*{4}}
\put(0,0){\line(0,1){40}}
\put(0,0){\line(2,1){40}}
\put(0,40){\line(2,-1){40}}
\put(0,0){\vector(2,1){22.5}}
\put(40,20){\vector(-2,1){22.5}}
\put(0,40){\vector(0,-1){22.5}}
\end{picture}}

\put(240,40){\begin{picture}(40,40)
\put(0,0){\color{blue} \circle*{4}}
\put(0,40){\color{green} \circle*{4}}
\put(0,0){\line(0,1){40}}
\put(0,0){\line(2,1){40}}
\put(0,40){\line(2,-1){40}}
\put(0,0){\vector(2,1){22.5}}
\put(40,20){\vector(-2,1){22.5}}
\put(0,40){\vector(0,-1){22.5}}
\end{picture}}

\put(240,80){\begin{picture}(40,40)
\put(0,40){\color{red} \circle*{4}}
\put(0,0){\line(0,1){40}}
\put(0,0){\line(2,1){40}}
\put(0,40){\line(2,-1){40}}
\put(0,0){\vector(2,1){22.5}}
\put(40,20){\vector(-2,1){22.5}}
\put(0,40){\vector(0,-1){22.5}}
\end{picture}}

\put(200,60){\begin{picture}(40,40)
\put(0,0){\color{red} \circle*{4}}
\put(0,40){\color{blue} \circle*{4}}
\put(0,0){\line(0,1){40}}
\put(0,0){\line(2,1){40}}
\put(0,40){\line(2,-1){40}}
\put(0,0){\vector(2,1){22.5}}
\put(40,20){\vector(-2,1){22.5}}
\put(0,40){\vector(0,-1){22.5}}
\end{picture}}

\put(160,20){\line(4,1){80}}
\put(160,20){\vector(4,1){62.5}}

\put(240,120){\line(-4,1){80}}
\put(240,120){\vector(-4,1){22.5}}

\put(120,0){\color{red} \circle*{4}}
\put(200,0){\color{blue} \circle*{4}}
\put(160,20){\color{green} \circle*{4}}
\put(160,40){\color{green} \circle*{4}}
\put(160,120){\color{green} \circle*{4}}
\put(160,140){\color{green} \circle*{4}}
\put(120,160){\color{blue} \circle*{4}}
\put(200,160){\color{red} \circle*{4}}

\put(200,0){\line(-1,0){80}}
\put(200,0){\vector(-1,0){42.5}}
\put(120,0){\line(2,1){40}}
\put(120,0){\line(1,1){40}}
\put(120,0){\vector(2,1){21.5}}
\put(120,0){\vector(1,1){21.5}}
\put(160,20){\line(2,-1){40}}
\put(160,20){\vector(2,-1){21.5}}
\put(160,40){\line(1,-1){40}}
\put(160,40){\vector(1,-1){21.5}}
\put(200,100){\line(-1,0){80}}
\put(200,100){\vector(-1,0){42.5}}
\put(120,100){\line(2,1){40}}
\put(120,100){\line(1,1){40}}
\put(120,100){\vector(2,1){21.5}}
\put(120,100){\vector(1,1){21.5}}
\put(160,120){\line(2,-1){40}}
\put(160,120){\vector(2,-1){21.5}}
\put(160,140){\line(1,-1){40}}
\put(160,140){\vector(1,-1){21.5}}

\put(120,160){\line(1,0){80}}
\put(120,160){\vector(1,0){42.5}}
\put(200,160){\line(-2,-1){40}}
\put(200,160){\line(-1,-1){40}}
\put(200,160){\vector(-2,-1){21.5}}
\put(200,160){\vector(-1,-1){21.5}}
\put(160,140){\line(-2,1){40}}
\put(160,140){\vector(-2,1){21.5}}
\put(160,120){\line(-1,1){40}}
\put(160,120){\vector(-1,1){21.5}}

\put(120,60){\line(1,0){80}}
\put(120,60){\vector(1,0){42.5}}
\put(200,60){\line(-2,-1){40}}
\put(200,60){\line(-1,-1){40}}
\put(200,60){\vector(-2,-1){21.5}}
\put(200,60){\vector(-1,-1){21.5}}

\put(160,40){\line(-2,1){40}}
\put(160,40){\vector(-2,1){21.5}}
\put(160,20){\line(-1,1){40}}
\put(160,20){\vector(-1,1){21.5}}

\put(120,0){\line(0,1){160}}
\put(200,0){\line(0,1){160}}
\put(120,60){\vector(0,1){22.5}}
\put(120,60){\vector(0,-1){22.5}}
\put(120,160){\vector(0,-1){42.5}}
\put(200,100){\vector(0,-1){22.5}}
\put(200,100){\vector(0,1){22.5}}
\put(200,0){\vector(0,1){42.5}}
\put(200,60){\line(-2,3){40}}
\put(200,60){\vector(-2,3){22.5}}
\put(160,120){\line(-2,-3){40}}
\put(160,120){\vector(-2,-3){22.5}}
\put(120,100){\line(2,-3){40}}
\put(120,100){\vector(2,-3){22.5}}
\put(160,40){\line(2,3){40}}
\put(160,40){\vector(2,3){22.5}}
\put(120,60){\line(-2,-1){40}}
\put(120,60){\vector(-2,-1){22.5}}
\put(80,120){\line(2,-1){40}}
\put(80,120){\vector(2,-1){22.5}}
\put(240,40){\line(-2,1){40}}
\put(240,40){\vector(-2,1){22.5}}
\put(200,100){\line(2,1){40}}
\put(200,100){\vector(2,1){22.5}}
\put(-10,80){\makebox(0,0){0}}
\put(330,80){\makebox(0,0){21}}
\put(40,50){\makebox(0,0){1}}
\put(40,110){\makebox(0,0){2}}
\put(69,80){\makebox(0,0){3}}
\put(80,30){\makebox(0,0){5}}
\put(80,130){\makebox(0,0){4}}
\put(120,-10){\makebox(0,0){11}}
\put(120,170){\makebox(0,0){10}}
\put(200,-10){\makebox(0,0){13}}
\put(200,170){\makebox(0,0){14}}
\put(240,30){\makebox(0,0){19}}
\put(240,130){\makebox(0,0){20}}
\put(280,50){\makebox(0,0){23}}
\put(280,110){\makebox(0,0){22}}
\put(160,13){\makebox(0,0){6}}
\put(160,32){\makebox(0,0){12}}
\put(160,128){\makebox(0,0){15}}
\put(160,147){\makebox(0,0){9}}
\put(110,60){\makebox(0,0){7}}
\put(110,100){\makebox(0,0){8}}
\put(212,60){\makebox(0,0){17}}
\put(212,100){\makebox(0,0){16}}
\put(252,80){\makebox(0,0){18}}
\put(0,80){\circle{7}}
\put(320,80){\circle{7}}
\put(-5,85){$\ast$}
\end{picture}
\end{center}
\caption{The ${\cal E}_{21}$ generalized Dynkin diagram}
\end{figure}
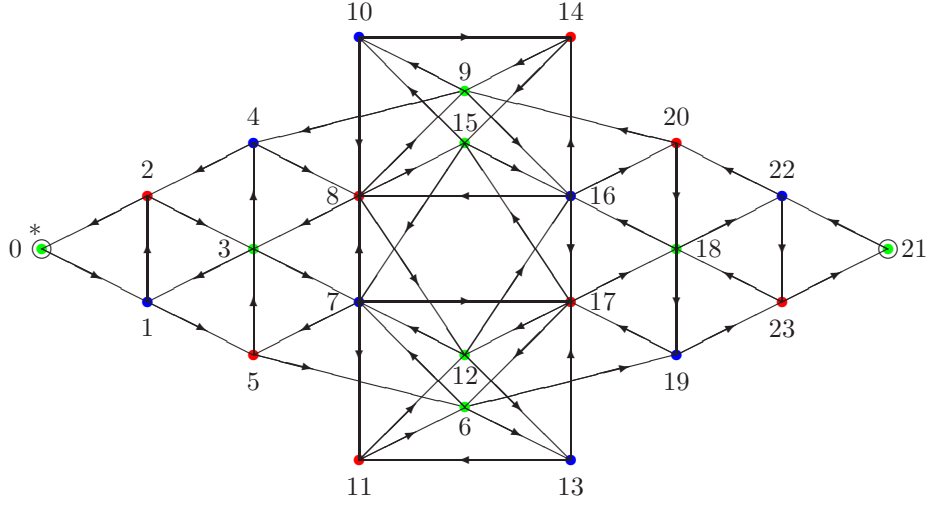

\subsubsection{Restriction and induction mechanism}
\begin{figure}
\mbox{\scalebox{0.9}{\includegraphics{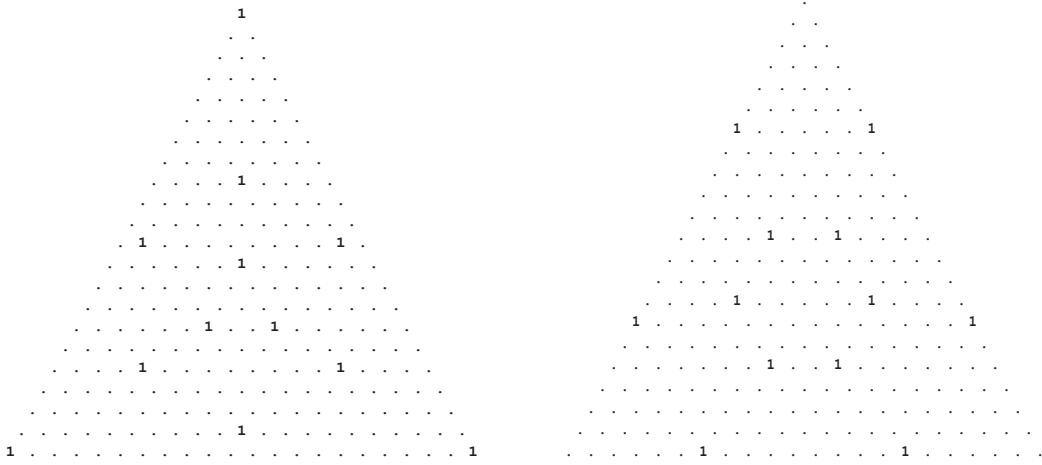}}}
\caption{Induction corresponding to the two extreme points of ${\cal E}_{21}$}
\label{fig:quantuminvariants21}
\end{figure}

The easiest method is to determine first the fused matrices $F_{i}$
by using the recurrence formula for $su(3)$. Essential matrices
$E_{a}$ -- and in particular $E_{0}$ -- are then obtained in the
usual way from the $F_{i}$'s. The first column of $E_{0}$ gives the
quantum invariants, it is displayed on the left array of Figure \ref{fig:quantuminvariants21}.

One can check that the values of the modular operator $T$, calculated
for ${\cal A}_{21}$, are equal
for all non-zero entries of this table.
The same property is also true for the column of $E_{0}$ associated with the
rightmost point of the ${\cal E}_{21}$ graph (right array of Figure \ref{fig:quantuminvariants21}).
However, $T$, when evaluated on non-zero entries of the $22$ other
columns of $E_{0}$, is not constant. We conclude that the set $J$ 
charactering the
ambichiral points of $Oc({\cal E}_{21})$ is a set with two
elements: the two extreme vertices of ${\cal E}_{21}$. The values of
the modular exponent $\hat{T}$ obtained for these two points are 
$\hat{T}=21$ and $\hat{T}=39$.

The dimensions $d_{j}$, with $j = (\lambda_{1}, \lambda_{2})$
of the $253$ blocks of the bialgebra ${\cal B}{\cal E}_{21}$, for the first law determined
by composition of  endomorphisms, are
obtained by summing matrix elements of matrices $F_{j}$. We obtain:
$dim({\cal B}{\cal E}_{21}) = \sum_{j} d_{j}^{2} = 480701952 = (2)^{9} (3)^{4} (67)^{1} (173)^{1}$, 
and also $\sum_{j} d_{j} = 288576 = (2)^{6} (3)^{3} (167)^{1}$.

\begin{figure}
\begin{center}
%\mbox{\rotatebox{0.5}{\scalebox{0.9}{\includegraphics{DimPathsE21.eps}}}}
\mbox{\scalebox{0.9}{\includegraphics{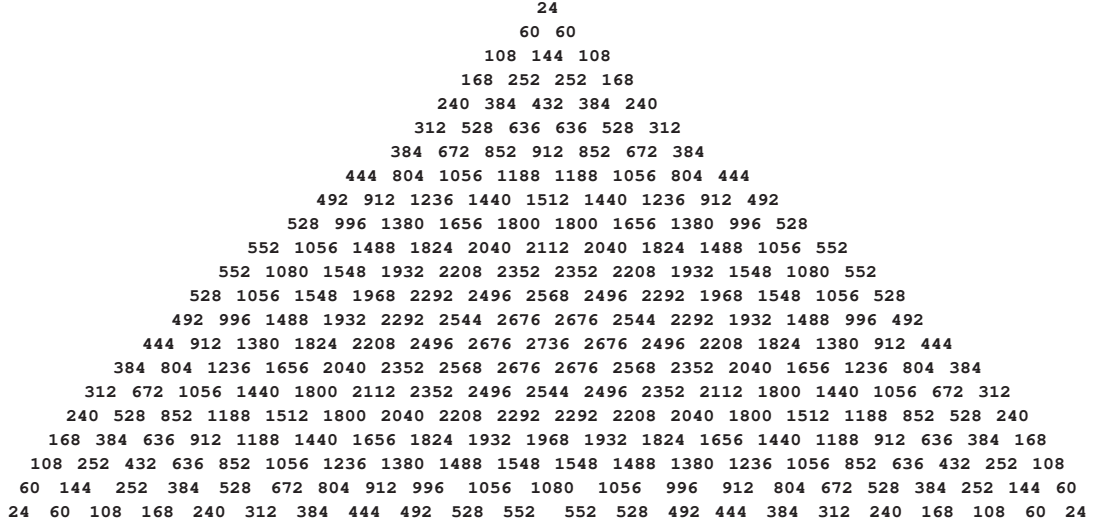}}}
\caption{Dimensions of spaces of blocks (law $\circ$)}
\label{fig:dimofblockspaths}
\end{center}
\end{figure}

\subsubsection{Determination of the graph algebra and of matrices
$G_{a}$}
The determination of graph matrices $G_{a}$ comes from the graph 
${\cal E}_{21}$ itself. To ease the calculation, it is
worth noticing that graph matrices associated with points symmetric
with respect to the horizontal symmetry axis of the graph are
transposed. We have for example $G_5=G_1.G_1-G_2$ 
so that $G_4=(G_5)^{t} = G_2.G_2 - G_1$.
Their determination is straightforward from vertices $1$ to $9$.
We then use the fact that $G_6.G_{21}=G_9$ to compute the matrix $G_{21}$
associated with the rightmost point of the graph. 
Multiplying a vertex $k$ by the vertex $21$ gives a vertex which is the symmetric  of $k$
with respect to the center of the graph (the center of a star).
In graph algebra terms, we get for example $G_{5}.G_{21}=G_{20}$ and 
$G_8.G_{21}=G_{17}$. It is then easy to compute the matrices associated 
with all the other vertices of the graph. 
The most important result, for what follows, is that $G_{21}.
G_{21} = G_{0}$.

\subsubsection{Quantum symmetries}
As already discussed, the subspace $J$ of ${\cal E}_{21}$
determining the algebra of quantum symmetries  is spanned by $0$
and $21$; we set $Oc({\cal E}_{21}) = {\cal E}_{21}\otimes_{J}{\cal E}_{21}$. 
This is a commutative algebra.
The left and right subalgebras $L$ and $R$ are
respectively spanned by $a \otimesdot 0 $ and by
$0 \otimesdot a$, where $a = 0,1,\ldots 23$.
Both left and right chiral subgraphs have $24$ points.
The ambichiral subalgebra $A$, of dimension $2$ is spanned
by $ \{0 \otimesdot 0  = 21  \otimesdot
21  \}$ and by $\{ 0 \otimesdot 21  = 
21  \otimesdot
0 \}$. The supplementary subspace $C$ is spanned by
$u \otimesdot a$, where $u \in
\{1,2,3,4,5,6,7,8,10,11,12\}$ and $a$ takes all possible values (but
neither $0$ nor $21$). The total number of vertices of the Ocneanu
graph is therefore
$22 + 22 + 2 + 11 \times 22 = 288$, as expected from the naive
dimension count $24 \times 24/2 = 288$.
As usual, blocks corresponding to
the second structure of the bialgebra ${\cal B}{\cal E}_{21}$
are labelled by the $288$ points of the Ocneanu
graph, and the dimension $d_{x}$ of the block $x$
is obtained by summing the matrix elements
of matrices $S_{x} = G_{a} G_{b}$ when $x = a \otimesdot b$ runs over
the points of $Oc( {\cal E}_{21})$.
We find (subscript give multiplicities of the blocks):

$$
\begin{array}{lcl}
\text{Ambichiral} &:& (24)_2 \\
\text{Left (not ambichiral)} &:& (60)_4 (108)_4 (132)_4 (144)_2 (168)_2 (216)_2 (252)_4 \\
\text{Right (not ambichiral)}&:& (60)_4 (108)_4 (132)_4 (144)_2 (168)_2 (216)_2 (252)_4 \\
\text{Supplement} &:& 
(168)_{8} (312)_{16} (384)_{16} (420)_{8} (492)_{8} (600)_{8} (636)_{8} (744)_{32} (804)_{8} (936)_{8} \\
{} & {} & (948)_{8} (996)_{8} (1080)_{2} (1188)_{8} (1236)_{8} 
(1272)_{4} (1440)_{16} (1512)_{2} (1548)_{8} \\
{} & {} & (1656)_{4} (1800)_{16} (1932)_{8} (1968)_{4} (2292)_{8} 
(2568)_{2} (2988)_{8} (3480)_{8}
\end{array}
$$

The quadratic and linear sum rules read:
$$
\begin{array}{cccclcl}
\sum d_x   &=& \sum d_i   &=& 288576 &=& (2)^6(3)^3(167)^1 \\ 
\sum d_x^2 &=& \sum d_i^2 &=& 480701952 &=& (2)^9(3)^4(67)^1(173)^1
\end{array}
$$

\subsubsection{Toric matrices and twisted partition functions}

We define the linear combination $U$ and $V$ of characters as follows:
$$
\begin{array}{ccc}
U  &=& \chi_{(0,0)}+\chi_{(0,21)}+\chi_{(1,10)}+\chi_{(4,4)}  +\chi_{(4,13)}+\chi_{(6,6)}  \\
{} &+& \chi_{(6,9)}+\chi_{(9,6)} +\chi_{(10,1)}+\chi_{(10,10)}+\chi_{(13,4)}+\chi_{(21,0)} \\
V  &=& \chi_{(0,6)}+\chi_{(0,15)}+\chi_{(4,7)} +\chi_{(4,10)} +\chi_{(6,0)} +\chi_{(6,15)} \\
{} &+& \chi_{(7,4)}+\chi_{(7,10)}+\chi_{(10,4)}+\chi_{(10,7)}+
         \chi_{(15,0)}+\chi_{(15,6)}
\end{array}
$$
The modular-invariant partition function ${\cal Z}_{{\cal E}_5}$ (associated with the vertex 
$0 \otimesdot 0$) and the
one associated with the vertex $0 \otimesdot 21$, that we call ${\cal Z}_{{\cal E}_5}^{'}$ are:

$$
\begin{array}{ccccl}
{\cal Z}_{{\cal E}_5} &\doteq& {\cal Z}_{0 \otimesdot 0} &=& U.\overline{U} + V.\overline{V} \\
{\cal Z}_{{\cal E}_5}^{'} &\doteq& {\cal Z}_{0 \otimesdot 21} &=& U.\overline{V} + V.\overline{U} 
\end{array}
$$
The first one agrees with the expression of \cite{Gannon}, the other, as explained
 in section 2.2, should be interpreted 
as a twisted partition function in a BCFT with one defect line \cite{PetZub:bcft}.
 There are $286$ other such functions
for the ${\cal E}_{21}$ diagram but these two are the only ones that are ambichiral.

\section{Appendices}

\subsection{About modular invariance}
The expressions for $S$ and $T$ can be taken from the theory of quantum groups
${\cal U}_{q}({\cal G})$ at roots of unity, \ie when $q = e^{\frac{i
\pi}{\mu \kappa}}$.
Here $\mu$ is half the length of a long root, so it is equal to $1$
when the Lie algebra ${\cal G}$ is  simply laced, which is the case in
particular for $su(2)$ and $su(3)$, and $\kappa$ is an arbitrary positive
integer, larger or equal to $h$, the dual Coxeter number of $\cal G$.
The equation $\kappa = h + k$ defines the ``level'' $k$.
One should consider a particular category whose objects are the
so-called tilting modules
of ${\cal U}_{q}({\cal G})$ and whose morphisms
are defined up to ``negligible morphisms'' (see for instance
\cite{Baki}); this is a semisimple ribbon and
modular category. This implies, in particular, that
a (projective) representation of $SL(2,\ZZ)$ can be
defined on the simple objects, thanks to two
matrices $s$ and $t$ and a phase $\zeta$ which are such that
$(st)^{3}= \zeta^{3} s^{2}$, $s^{2}= C$, $Ct=tC$ and $C^{2}=1$. The
matrix $C$ is called ``conjugation matrix'' and $t$ is the ``modular twist''.
For this category, $\zeta = e^{2 i \pi c/24}$ with $c =
(\kappa - h) dim({\cal G})/\kappa$.
The  expression for the $t$ matrix, in the case of an arbitrary Lie
algebra ${\cal G}$ is $ t_{mn} = \delta_{mn} q^{<<n, n + 2 \rho>>}$
where $\rho$ is half the sum of positive roots and $m$, $n$ are
elements of the weight lattice of  ${\cal G}$ characterizing the
representation $\tau_{m}$ and $\tau_{n}$. Here $<<.,.>>$ is an
invariant bilinear form
on $\cal G$ normalized by $<<\alpha,\alpha>>=2$ for a short root
$\alpha$. The corresponding general expression for the $s$ matrix is 
more involved
and we do not need it in our paper.
The same expressions for the modular generators can be
obtained from the Kac-Peterson formulae \cite{KP} for the modular
transformations of characters
of the affine Lie algebra $\widehat{\cal G}$, evaluated at the same
value $q = e^{\frac{i \pi}{\mu \kappa}}$.
Here $k = \kappa - h$ is indeed  equal to the usual level.

In the case of $su(2)$, the modular generators $S$, $T$, are as 
follows: $S=s$ and $T = t/ \zeta $.
The $SL(2,\ZZ)$ relations read then  $(ST)^3= S^2 = 1$, with $s_{mn} = 
\sqrt{\frac{2}{\kappa}}
\sin(\pi \frac{(m+1)(n+1)}{\kappa})$, for $0 \leq m,n \leq \kappa -2$ 
and $t_{mn} = e^{\frac{i \pi}{2\kappa} n(n+2)}
\delta_{mn}$.  Still for $su(2)$ we have
$c=3-6/\kappa$, so that $\zeta = e^{i \pi/4} e^{\frac{ -i \pi} 
{2\kappa}}$ and therefore
$$T_{mn} = exp[2i\pi(\frac{(n+1)^2}{4 \kappa} - \frac{1}{8})] 
\delta_{mn} $$
which is the expression used in the text.
One can explicitly see that the previous $SL(2,\ZZ)$ relations
hold.  It can be checked, from this expression that, $T^{8 \kappa} =1$ 
when $\kappa$ is odd and $T^{4 \kappa} = 1$ when $\kappa$ is even. 
This, by itself, is not enough to imply the following property, which
is nevertheless true, and was proven more than a hundred years ago
\cite{Hurwitz}: the above representation of $SL(2,\ZZ)$ factorizes
over the finite group $SL(2,\ZZ/8 \kappa \ZZ)$ when $\kappa$ is odd,
and factorizes over $SL(2,\ZZ/4 \kappa \ZZ)$ when $\kappa$ is even. 
So, in particular, $T^{40}=1$ for the $A_{4}$ graph ($40 = 8 \times
5$), but $T^{48}=1$ for the $A_{11}$ graph ($48 = 4 \times 12$).  In
the text, we use (for $su(2)$) a ``modular exponent'' defined by $\hat
T = (n+1)^{2}$ mod $4\kappa$, but it is clear that we could use as
well $n(n+2)$ mod $4\kappa$ or any other expression differing by a
constant shift.

\subsection{The general notion of essential paths on a graph $G$ of 
type ADE}

The following definitions are not needed if we only want to count the 
number of essential paths on a graph. They are necessary if we want to
obtain explicit expressions for them.
These definitions are adapted from \cite{Ocneanu:paths}, see also
several comments made in \cite{Coque:Qtetra} and \cite{Coque:Karpacz}.
Call $\beta$ the norm of the graph $G$ (the biggest eigenvalue of its 
adjacency matrix ${\cal G}$) 
and  $D_{i}$ the components of the (normalized) Perron Frobenius eigenvector.  
Call $\sigma_{i}$ the vertices of $G$ and, if 
$\sigma_{j}$ is a neighbour of 
$\sigma_{i}$, call $\xi_{ij}$ the oriented edge
from $\sigma_{i}$ to $\sigma_{j}$. If $G$ is unoriented (the case for $ADE$
and affine $ADE$ diagrams), each edge should be considered  as carrying
both orientations.
An elementary path can be written either as a finite 
sequence of consecutive (\ie neighbours on the graph) vertices, 
$[\sigma_{a_1} \sigma_{a_2} \sigma_{a_3} \ldots ]$,
or, better, as a sequence $(\xi(1)\xi(2)\ldots)$ of consecutive edges, with
$\xi(1) = \xi_{a_{1}a_{2}}= \sigma_{a_1} \sigma_{a_2} $,
$\xi(2) = \xi_{a_{2}a_{3}} = \sigma_{a_2}  \sigma_{a_3} $, \etc.
Vertices are considered as paths of length $0$.
The length of the (possibly backtracking) path $( \xi(1)\xi(2)\ldots 
\xi(p) )$ is $p$.
We call $r(\xi_{ij})=\sigma_{j}$, the range of $\xi_{ij}$
and $s(\xi_{ij})=\sigma_{i}$, the source of $\xi_{ij}$.
For all edges $\xi(n+1) = \xi_{ij}$ that appear in an elementary path, 
we set  ${\xi(n+1)}^{-1} \doteq \xi{ji}$.
For every integer $n >0$, the annihilation operator $C_{n}$,
acting on the vector space generated by elementary paths of length $p$ is defined
as follows:  if $p \leq n$, $C_{n}$ vanishes, whereas if $ p \geq  n+1$ then
$$
C_{n} (\xi(1)\xi(2)\ldots\xi(n)\xi(n+1)\ldots) = 
\sqrt\frac{D_{r(\xi(n))}}{D_{s(\xi(n))}} 
\delta_{\xi(n),{\xi(n+1)}^{-1}}
 (\xi(1)\xi(2)\ldots{\hat\xi(n)}{\hat\xi(n+1)}\ldots) 
$$

Here, the symbol ``hat'' ( like  in $\hat \xi$) denotes omission.
The result is therefore either $0$ or a linear combination of paths of length $p-2$.
Intuitively, $C_{n}$ chops the round trip that possibly appears
at positions $n$ and $n+1$.

A path is called essential if it belongs to 
the intersection of the kernels 
of the anihilators $C_{n}$'s.

Here comes an example of calculation for the  $E_{6}$ diagram (square brackets
enclose $q$-numbers),  
\begin{eqnarray*}
C_{3}(\xi_{01}\xi_{12}\xi_{23}\xi_{32}) &=&  \sqrt \frac {1}{[2]} 
(\xi_{01}\xi_{12}) \\
C_{3}(\xi_{01}\xi_{12}\xi_{25}\xi_{52}) &=&  \sqrt \frac {[2]}{[3]} 
(\xi_{01}\xi_{12})
\end{eqnarray*}

The following difference of non essential paths of length $4$ starting
at $\sigma_{0}$ and ending at $\sigma_{2}$ is an essential path of
length $4$ on $E_{6}$: $$\sqrt{[2]} (\xi_{01}\xi_{12}\xi_{23}\xi_{32})
- \sqrt \frac{[3]}{[2]} (\xi_{01}\xi_{12}\xi_{25}\xi_{52})
= \sqrt{[2]}  [0,1,2,3,2]- \sqrt \frac{[3]}{[2]}[0,1,2,5,2]$$
Remember the values of the $q$-numbers: $[2] = \frac{\sqrt 2}{\sqrt 3
-1}$ and $[3] = \frac{2}{\sqrt 3 -1}$.

Acting on elementary path of length $p$, the creating operators
$C^{\dag}_{n}$ are defined as follows: if $n > p+1$, $C^{\dag}_{n}$
vanishes and, if $n \leq p+1$ then, setting $j = r(\xi(n-1))$,
$$
C^{\dag}_{n} (\xi(1)\ldots\xi(n-1)\ldots) = \sum_{d(j,k)=1}
\sqrt(\frac{D_{k}}{D_{j}})  (\xi(1)\ldots\xi(n-1)\xi_{jk}\xi_{kj}\ldots)
$$
The above sum is taken over the neighbours $\sigma_{k}$ of $\sigma_{j}$ on the graph.
Intuitively, this operator adds one  (or several) small round trip(s) 
at position $n$. 
The result is therefore either $0$ or a linear combination of paths of 
length $p+2$.
For instance, on paths of length zero (\ie vertices),
$$
C^{\dag}_{1} (\sigma_{j}) = \sum_{d(j,k)=1}
\sqrt(\frac{D_{k}}{D_{j}}) \xi_{jk}\xi_{kj} = \sum_{d(j,k)=1}
\sqrt(\frac{D_{k}}{D_{j}}) \, [\sigma_{j}\sigma_{k}\sigma_{j}]
$$

Jones' projectors $e_{k}$ can be realized (as endomorphisms of 
$Path^p$) by 
$$
e_{k} \doteq \frac{1}{\beta} C^{\dag}_{k} C_{k} 
$$

The reader can check that all Jones-Temperley-Lieb relations 
between the $e_i$ are satisfied.
Essential paths can also be defined as elements of the intersection of the 
kernels of the Jones projectors $e_{i}$'s.

\subsection{The structure of $\mathcal{B}G$}

Paths on $G$ generate a vector space $Paths(G)$ which comes with a
grading: paths of homogeneous grade $j$ are associated with Young diagrams
of $SU(N)$.  In the case of $ su(2)$ this grading is just an integer
(to be thought of as a length or as a point of a diagram of type 
${\cal A}$).

What turns out to be most interesting is a particular vector subspace
$\mathcal{E} = EssPaths(G)$ of $Paths$ whose elements are called
``essential paths'' (see above definition). This subspace is is itself
graded in the same way as $Paths$.

We then consider the graded algebra of endomorphisms of
essential paths $$\mathcal{B}G  =  End_{\sharp}(EssPaths) =
\bigoplus_{j=0,r-1} End(EssPaths^{j})$$ which, by definition, is an 
associative algebra.
By using the fact that paths on the chosen diagram can be
concatenated, one may define \cite{Ocneanu:paths} {\sl another}
multiplicative associative structure on $\mathcal{B}G$ that we call
convolution product (see our comments in the next subsection).
This vector space with two algebra structures
is  called, by A. Ocneanu, the ``Algebra of
double triangles".

Existence of a scalar product allows one to transmute one of the
multiplications (for instance the convolution product) into a
co-multiplication and it happens that the coproduct $\Delta$ is
compatible with the product (in the sense that we have the
homomorphism property $\Delta (u v) = \Delta u \otimes \Delta v$). 
$\mathcal{B}G$ is therefore a bialgebra.
However, $\mathcal{B}G$ is not a Hopf algebra but a weak Hopf algebra
(or quantum groupoid).  
 This statement should be taken with a grain of salt: see our comments in the next subsection.
 General axioms for weak Hopf algebras are 
given in \cite{Sz}. In the present case,  the following axiom for Hopf algebras
fails to be satisfied: the coproduct of the unit $\Delta \one = \one_{1} \otimes \one_{2}$ is
not equal to $\one \otimes \one$ (as usual, a summation is
understood); several other axioms for
Hopf algebras are also modified: the counit is not an homomorphism
($\epsilon(xy)=\epsilon(x\one_{1})\epsilon(y\one_{2})$) and, if 
$\Delta^{2} x = x_{1} \otimes x_{2} \otimes x_{3}$, the 
compatibility axiom for the antipode is modified as follows $S(x_{1}) x_{2} 
\otimes x_{3} = \one_{1} \otimes x \one_{2}$.

\subsection{Remarks and open questions}

Essential paths for $ADE$ diagrams (\ie the $su(2)$ 
system) have been defined in several published papers but their  
analogues for higher systems (for instance the Di Francesco - Zuber 
diagrams), although  reasonably well understood by a few 
people, have never been described, as far as we know, in the 
litterature.

The general definition of the convolution product of $\mathcal{B}G$,
for $ADE$ diagrams, was given ``explicitly'' by A. Ocneanu in
\cite{Ocneanu:paths} by a rather difficult formula involving several
types of generalized quantum $6j$ symbols.  It is certainly interesting to
know this general formula, but, in our opinion, this
expression is not very helpful for a practical investigation of the
different cases.

 The fact that $\mathcal{B}G$ is a weak Hopf algebra is a claim 
 that belongs to the folklore,  but we are not aware of any general reference showing 
 that all the axioms of \cite{Sz} are indeed verified in this situation.
 The authors (together with A. Garcia and R. Trinchero) have however 
 checked that it is so in a number of particular cases belonging to the ADE series
  and are working on a general proof.

   Another possibility for defining the convolution product of
  $\mathcal{B}G$ is to make use of the notion of cell systems.  This
  general notion was defined in
  \cite{Ocneanu:paragroups}; it is also described in \cite{EvansKawa:book}
  and it is used, in a particular context  by  \cite{Roche:Oc}.
  We cannot summarize this theory here. 
  Let us just mention that a cell system involves four graphs (top,
  bottom, left and right) with matching properties and that, in the
  present case, the top and bottom graphs are the same $ADE$ diagram
  $G$.  Cells are rectangles with top and bottom edges which are also
  edges of the given graph(s).  Macrocells have top and bottom edges
  (or ``horizontal paths'') that co\" \i ncide with the essential paths
  on $G$; their left and right edges are called ``vertical paths''. 
To every cell system one can associate ```connections'' which are particular maps
associating complex numbers with  cells  or macrocells.
  These numbers, in turn, can be used to define the structure constants of the
  algebra we are looking for.  For every point of the graph $Oc(G)$ there is an
  irreducible connection on the cell system (or an irreducible quantum symmetry).
  Although it  seems to provide (at the time of this writing) the shortest road to the
  explicit construction of the bialgebra $\mathcal{B}G$,  this construction is unfortunately not
  explicitely available in the litterature.

 Among other results, and in the framework of statistical mechanics,
 the paper \cite{PetZub:Oc} gives many useful relations between the
 vertical product of  $\mathcal{B}G$ (the product of
 endomorphisms acting on $EssPaths(G)$) and its horizontal product (or
 convolution product).  There are indeed several families of numeral
 constants that appear as structure constants for these two products,
 or that appear as coefficients of a kind of Fourier transform relating the two. 
 These constants look like generalized quantum 6j symbols and obey different
 types of (mixed) pentagon equations which themselves generalize the
 quantum group version of the Biedenharn - Elliot identity.  As
 discussed in \cite{Sz}, any solution of this ``Big Pentagon
 Equation'' (involving six different types of generalized $6j$
 symbols) determines the structural maps of a weak $C^{*}$ Hopf
 algebra.  Unfortunately, we do not know a single reference that
 describes a practical implementation 
  of this general construction (and gives the values of these structure constants)
  for  the bialgebras $\mathcal{B}G$ associated with specific $ADE$ diagrams
 or with their higher generalizations.

Graphs $Oc(G)$, encoding the structure of the algebra of quantum
symmetries of the diagram $G$, have been ``conceptually'' defined by
A. Ocneanu in terms of the block structure of $\mathcal{B}G$ for
its convolution product, but it is interesting to notice that, to our knowledge, they
were never obtained in this way\ldots
Clearly, it would be interesting to do so.  We repeat that our modest
purpose, in the present paper, was to observe that known 
Ocneanu graphs (or algebras), in the ADE cases,  could be recovered, in most cases,  from
the modular properties of the $T$ matrix; we then used
this observation to study several cases belonging to the $su(3)$ system.
The problem of  deducing Ocneanu graphs from the explicit structure of
the bialgebra  $\mathcal{B}G$, in the different cases, is a much more difficult and interesting program that it would be nice to investigate.

\bigskip

Here comes a short list of open questions that, we hope, may trigger 
the interest of the reader:

\begin{itemize}
\item Give a simple definition -- valid in all cases -- of the 
convolution product of $\mathcal{B}G$.
\item Show that this bialgebra is indeed a weak Hopf algebra in all 
cases.
\item Is it possible to find a kind of multiplication on $EssPaths(G)$ 
that would allow one to construct $\mathcal{B}G$ in a functorial (and simple)  way ?
\item Determine explicitly the graphs $Oc(G)$ directly from the study  of the corresponding bialgebra $\mathcal{B}G$.
\item Find a simple algorithm allowing one to calculate all irreducible connections on cell systems (\ie the values of cells) in all $ADE$ or generalized $ADE$ cases.
\item Precise the relation (if any) between the generalized Coxeter-Dynkin 
systems and the finite subgroups of Lie groups.
\item What is the interpretation of all these contructions in terms of
the finite dimensional Hopf quotients of $U_{q}(SL2)$ at roots of unity ?
\item Can one, in some sense, ``supersymmetrize''  these 
constructions ?
\item What is the origin of the linear sum rules ?
\item What is the origin of the quantum sum rules ?
\item As we know, toric matrices (twisted or not) described in the text can be
interpreted as partition functions (with or without defect lines) on a
torus, at the critical point, for affine models (WZW models).  Clearly
this framework can be generalized in several directions: one may
consider more general correlation functions, replace affine models by
(generalized) minimal models, replace the torus by higher genus
surfaces\ldots
\item We know explicitly how to generalize the $ADE$ diagrams in the cases of
$su(3)$ and $su(4)$ and a definition of what are the ``generalized Coxeter-Dynkin systems''
 was briefly mentionned in \cite{Ocneanu:MSRI} but  a detailed description of this notion is clearly needed.
\item What kind of algebraic structures (generalizing the notion of Lie 
algebras) can one associate with a diagram belonging to such a generalized system ?
\end{itemize}

\nopagebreak[1]

%\vskip .5cm

%\noindent{\bf\Large 6 \; Acknowledgements}

%\vskip .5cm

\section*{Acknowledgments}

We thank the referee for his careful reading of the manuscript and for 
his questions and comments.\\
One of us (R.C.) wants to thank the Centro Brasileiro de Pesquisas
F\'{\i}sicas (CBPF, Rio de Janeiro), where
part of this work was done, for its hospitality.\\
G. Schieber would like to thank the Conselho Nacional de Desenvolvimento Cient\'{\i}fico
e Tecnol\'ogico, CNPq, and the Coordena\c{c}\~ao de Aperfei\c{c}oamento de Pessoal de N\'{\i}vel
Superior, CAPES, Brazilian Research Agencies, for financial support.

\end{document}